\documentclass[12pt]{iopart}
\usepackage{iopams}
\usepackage{amssymb,graphicx,braket}
\usepackage{hyperref}
\usepackage{multirow}
\usepackage{ulem}
\usepackage{comment}
\usepackage{xcolor}
\usepackage{hyperref}
\usepackage{xcolor}

\newcommand{\beq}{\begin{equation}}
\newcommand{\eeq}{\end{equation}}
\newcommand{\beqa}{\begin{eqnarray}}
\newcommand{\eeqa}{\end{eqnarray}}

\newcommand{\eqref}[1]{(\ref{#1})}
\newcommand{\appropto}{\mathrel{\vcenter{
  \offinterlineskip\halign{\hfil$##$\cr
    \propto\cr\noalign{\kern2pt}\sim\cr\noalign{\kern-2pt}}}}}
    
\begin{document}

\title{Regressions on quantum neural networks at maximal expressivity}

\author{I.~Panadero$^{1,2,3}$, Y. Ban$^{1,4}$, H.~Espin\'os$^{1}$, R.~Puebla$^{1}$, J.~Casanova$^{3, 5, 6}$ E.~Torrontegui$^{1}$}
\address{$^{1}$ Departamento de F\'isica, Universidad Carlos III de Madrid, Avda. de la Universidad 30, 28911 Legan\'es, Spain}
\address{$^{2}$ Arquimea Research Center, Camino las Mantecas s/n, 38320 Santa Cruz de Tenerife, Spain}
\address{$^{3}$ Department of Physical Chemistry, University of the Basque Country UPV/EHU, Apartado 644, 48080 Bilbao, Spain}
\address{$^{4}$ TECNALIA, Basque Research and Technology Alliance (BRTA), 48160 Derio, Spain}
\address{$^{5}$ EHU Quantum Center, University of the Basque Country UPV/EHU, Leioa, Spain}
\address{$^{6}$ IKERBASQUE, Basque Foundation for Science, Plaza Euskadi 5, 48009 Bilbao, Spain}
\ead{ipanadero@gmail.com / eriktorrontegui@gmail.com}

\begin{abstract}
We analyze the expressivity of a universal deep neural network that can be organized as a series of nested qubit rotations, accomplished by adjustable data re-uploads. While the maximal expressive power increases with the depth of the network and the number of qubits, it is fundamentally bounded by the data encoding mechanism. Focusing on regression problems, we systematically investigate the expressivity limits for different measurements and architectures. The presence of entanglement, either by entangling layers or global measurements, saturate towards this bound. In these cases, entanglement leads to an enhancement of the approximation capabilities of the network compared to local readouts of the individual qubits in non-entangling networks. We attribute this enhancement to a larger survival set of Fourier harmonics when decomposing the output signal. 
\end{abstract}

\vspace{2pc}

\section{Introduction}\label{Introduction}
The convergence of machine learning (ML) and quantum computing leads to the birth of quantum machine learning \cite{Kak1995, Biamonte2017, Schuld2019}, which exploits both information processing paradigms to provide better algorithms. This is achieved either by using ML to improve the efficiency of quantum protocols \cite{Carrasquilla2017, Deng2017, Torlai2018, Fosel2018, Aharon2019} or by taking advantage of quantum resources to create new paradigms with better performance \cite{Gupta01, Papapro14, Benedetti17, Sentis15, Havlicek19}, such as quantum neural networks (QNN). While there are several studies comparing the expresssivity of classical and quantum artificial neural networks \cite{Abbas2021, Qian2021}, the goal of beating classical ML is undergoing critical revision \cite{Schuld2022}. Recent works, in the quest of finding better quantum models for ML purposes, characterize and compare the expressivity among different networks typologies \cite{Funcke2021, Beer2021, Gratsea2022, Wilkinson2022,Casas2023}. This is made possible by the emergence of unitary-based QNNs \cite{Cao2017, Farhi2018, Torrontegui2019, PerezSalinas2020, Mangini2020, Ban2021a, Ban2023}, which offer experimental feasibility across multiple quantum platforms, such as trapped ions \cite{Huber2021, Dutta2021} and superconducting circuits \cite{PerezSalinas2020,Havlicek19b, Pechal2021,Moreira2022}. 

On the one hand, a perceptron-type QNN, that can be implemented as a quasi-adiabatic passage in an Ising model \cite{Torrontegui2019, Ban2021a, Ban2023}, expresses any continuous function as a weighted linear combination of different adjustable sigmoids based on the universal approximation theorem \cite{Cybenko1989}. Correlations resulting from the entanglement among various perceptrons provide these artificial networks with at least as much power as their classical counterparts. However, the entanglement is intrinsically generated along hidden layers, and understanding its role in improving the network approximation capability in classification and regression problems remains a challenge. On the other hand, QNNs based on parametric quantum circuits, by virtue of the Solovay–Kitaev theorem \cite{Nielsen2010}, are universal classifiers, and are implemented in repeat-until-success \cite{Cao2017}, random circuits \cite{Farhi2018,Benedetti2019}, or by data re-uploading \cite{PerezSalinas2020}. A single qubit has already been employed to approximate continuous functions, specifically in the context of regression problems \cite{PerezSalinas2021}. Quantum resources may be introduced in an easily controllable manner by adding entangling gates into the topology of the network. 

While the expressivity of all these unitary-based QNN models is fundamentally bounded by the data encoding strategy \cite{schuld2021effect}, we show that the maximal performance in regression tasks is conditioned by the final QNN state readout. To this end, we assess the expressivity of a universal deep QNN that follows a data re-uploading strategy, employing various measurements and architectures that vary in terms of network depth and number of qubits. This construction mitigates the discrepancies attributed to the use of different models, enabling a fair comparison of networks with the same topology, but distinguished by the presence of quantum resources, such as entanglement. We make use of a partial Fourier series representation of the QNN outputs to characterize and bound the maximal expression power of the network. In particular, we find that the maximal performance is conditioned by the final QNN state readout. With the aid of the teacher-student scheme \cite{Gratsea2022}, we systematically benchmark the network accuracy in regression problems, showing that global measurements of the multiqubit system or the use of a final entangling layer leads to the saturation of the maximal expressive bound, optimizing the performance of the network.

The article is structured as follows. We begin by introducing a universal quantum regressor based on re-uploading continuous input data, which results in a trainable QNN. Then, we quantify the QNN's expressivity through partial Fourier decomposition and numerically benchmark several QNN distinguished by the presence of entanglement. The manuscript ends with the main conclusions and prospective research lines.  

\section{Deep data re-uploading QNN}

We propose an encoding scheme of arbitrary normalized real functions $f(\vec x)$ $\in$ $[0,1]$ through the degrees of freedom of a set of qubits. These qubits are assumed to constitute a circuit that performs a unitary transformation of the initial state $\ket{\phi_0}$ into the final output state 
\beq
\label{single}
\ket{\phi(\vec x)}=\hat{\mathcal{U}}(\vec x)\ket{\phi_0},
\eeq
by the unitary transformer $\hat{\mathcal{U}}(\vec x)$, where $\vec x =( x_1,x_2, \dots,x_n)\in \mathbb{R}^n$ are independent input variables that set the topology of the circuit. The circuit is approximated by a data re-uploading QNN with an adjustable set of parameters $\vec\theta=\{\vec\theta_1,\vec\theta_2,\cdots\vec\theta_L\}$, 
\beq
\hat{\mathcal{U}}(\vec x)\simeq\hat U(\vec{x},\vec\theta)=\prod_{l=1}^L\hat{ \mathcal{L}}_l(\vec{x},\vec\theta_l),
\eeq
where we adopt the convention $\prod_{n=1}^NA_n=A_N\cdot A_{N-1}...\cdot A_1$, with the dot being the standard matrix multiplication. The $l$-th layer is represented by the unitary transformation 
 $\hat{ \mathcal{L}}_l(\vec{x},\vec\theta_l)$, with input data $\vec{x}$ and undetermined parameters $\vec\theta_l$. 
Similar to perceptron-based neural networks \cite{Torrontegui2019}, the complexity (or interconectivity) of the network increases with the number of layers $L$. We choose the unitary transformation for each layer to be constituted by a single qubit SU(2) rotation with the data encoded as the rotation angle,
\beq
\hat{\mathcal{L}}_l(\vec x,\vec\theta_l) = \hat R_x(2(\vec\omega_l\cdot\vec x+\beta_l)) \hat R_y(2\alpha_l),
\label{eqn:layer}
\eeq
where $\hat R_\gamma(2\tau)=e^{i\tau\hat\sigma_\gamma}$ is the rotation around the $\gamma-$axis, with $\gamma\equiv\{x,y,z\}$ and $\hat\sigma_{\gamma}$ representing the standard Pauli matrices. For every layer, there are $n+2$ adjustable parameters $\vec\theta_l=\{\vec \omega_l,\beta_l, \alpha_l \}\in\{\mathbb{R}^n,\mathbb{R},\mathbb{R}\}$, and the scalar product is defined as $\vec\omega_l\cdot\vec x=\omega_{l,1}x_1+\dots+\omega_{l,n}x_n$. Thus, the complexity of the circuit increases linearly with the dimension of the input data. 

The output of the QNN, $F(\vec{x}, \vec\theta)$, is used to approximate a target function $f(\vec{x})$, and it is obtained by calculating the expectation value of an observable $\hat{\mathcal{M}}$ in the output state $\ket{\phi(x)}$,
\beq
\label{eqn:map}
F(\vec x, \vec\theta) = |\bra{\phi(\vec x)}\hat \mathcal{  M}\ket{\phi(\vec x)}| \rightarrow{f(\vec{x})}.
\eeq
Initially, we choose $ \hat \mathcal{  M}=\hat\sigma_z$, such that the mapping is done in the excitation amplitude of the qubit. However, more sophisticated mappings are possible, e.g., $ \hat \mathcal{ M} = \hat O = u_o \mathbb{I} + u_x \hat \sigma_x + u_y \hat \sigma_y + u_z \hat \sigma_z$, with $u_o, u_x, u_y, u_z \in \mathbb{R}$ also trainable parameters. The optimal network topology is constituted by the set of parameters $\vec\theta$ that minimizes a particular cost function $\mathcal{C}$, i.e.,  $\vec\theta_T=\mbox{argmin}_{\vec\theta}\ \mathcal{C}(f; \vec\theta)$. For the considered regression problems, we use the squared error loss function, which corresponds to
\beq
\mathcal{C}(f; \vec\theta) =\frac{1}{K}\sum_{k=1}^{K}\bigg[F(\vec x^k,\vec\theta) -f(\vec x^k)\bigg]^2,
\label{eqn:cost}
\eeq
where $K$ is the total input samples of $\vec x$. The optimal parameters $\vec\theta_T$ for a particular target function are obtained minimizing Eq. (\ref{eqn:cost}) using gradient descent optimization, with $\vec\theta$ initialized randomly. To explore a larger region of the parameter landscape, the minimization is repeated $R$ times.

Note that the topology given by Eq. (\ref{eqn:layer}) is able to connect any two states on the Bloch sphere. However, in order to capture non-trivial patterns in the original data $\vec{x}$ and to approximate an arbitrary real function $f(\vec x)$, this data $\vec{x}$  needs to be reintroduced in the subsequent layers \cite{PerezSalinas2020}. 
Meanwhile, the complexity and approximation power of the network can be increased by extending the model from one qubit to an arbitrary number of qubits, $Q$. We first consider that the qubits forming the QNN do not interact. Thus, the total transformation is the outer product of the unitary transformations of all qubits after each layer, such that
\beq
\hat U (\vec{x},\vec\theta) =\prod_{l=1}^L \bigotimes_{q=1}^{Q} \hat{ \mathcal{L}}_{l,q} (\vec x, \vec\theta_{l,q}), 
\label{eqn:CNN}
\eeq
where each qubit has a set of adjustable parameters $\vec\theta_{l,q} = \{ \vec \omega_{l,q}, \beta_{l,q}, \alpha_{l,q} \}$ and the approximation of $f(\vec x)$ is done via $ \hat \mathcal{ M}=\hat \sigma_z^{tot}$, with $\hat \sigma_z^{tot}=\sum_{q=1}^Q\hat\sigma_z^q$. 
The QNN has a total number $N_p = LQ(n+2)$ of adjustable parameters $\vec\theta$. We now introduce entangling gates at each layer among the different qubits in the QNN model described by Eq. (\ref{eqn:CNN}),
\begin{equation}
\label{eqn:QNN}
\hat U(\vec{x},\vec\theta) = \prod_{l=1}^L  \hat E^l\cdot\bigg(  \bigotimes_{q=1}^{Q}\hat{\mathcal{L}}_{l,q} (\vec x, \vec\theta_{l,q})\bigg).
\end{equation}
Although there are different ways of introducing entanglement to enrich the network topology (by modifying either the entangling gates at each layer or the entanglement allocation), for simplicity, we consider here a fixed gate applied at each layer. In particular, to emphasize the role of the entanglement, we define $\hat E^l$ as a maximally entangling gate, which produces a generalization of Bell states when applied to the computational basis \cite{Aryeh2014},
\begin{equation}
\label{eqn:en1}
\hat E^l = \prod_{i=1}^{Q-1}\bigg((\bigotimes_{i-1} \mathbb{I}) \otimes \hat E_2\otimes ( \bigotimes_{Q-i-1}\mathbb{I})\bigg),
\end{equation}
with  
%
\begin{equation}
\label{eqn:en2}
\hat E_2 = \frac{1}{\sqrt{2}}(\mathbb{I} \otimes \mathbb{I} + i \sigma_x \otimes \sigma_y).
\end{equation}
Note that the amount of entanglement is introduced as a fixed quantity, i.e., the gates in Eq. (\ref{eqn:en1}) do not introduce new adjustable parameters. To distinguish different architectures, we label the topology of the QNN as $\mathcal{A}_0(L,Q)$ ($\mathcal{A}_1(L,Q)$) with the subscripts $0$ ($1$) indicating the absence (presence) of entangling gates. Two architectures, $\mathcal{A}_0(L,Q)$ and $\mathcal{A}_1(L,Q)$, contain the same number $N_p$ of variational parameters $\vec\theta$ and only differ by the incorporation of entanglement. Consequently, we are able to evaluate exclusively the role of this quantum resource in the performance of the QNN.

\section{Approximation power}
\label{approx_pow}
In order to characterize the expressivity of the QNN, we use the Fourier decomposition of the unitary transformation produced by the QNN and compute the number of harmonics that do not vanish after encoding the output state through the observable $\hat{\mathcal{M}}$. 
We introduce a novel strategy for bounding the expressivity of the QNN \cite{schuld2021effect,casas2023multi} that enables to compute the Fourier amplitudes and to identify zero coefficients. 

For each qubit, the single layer unitary transformation given by Eq.~(\ref{eqn:layer}) is decomposed into the fundamental frequencies $\pm\vec\omega_{l,q}$ of the qubit,
\begin{eqnarray}
\fl
 \hat{\mathcal{L}}_{l, q}(\vec x, \vec\theta_{l,q}) = \hat R_x(2(\vec \omega_{l, q} \cdot\vec x + \beta_{l, q})) \hat R_y(2 \alpha_{l, q})
     =\hat T_{l, q}^+ e^{i \vec \omega_{l, q} \cdot \vec x} + \hat T_{l, q}^- e^{-i \vec \omega_{l, q} \cdot \vec x}, 
\label{eq:harmonic1q}
\end{eqnarray}
where 
\beq
    \hat T^+_{l, q} = \frac{1}{2} \left(\matrix{
    t_{l, q}^- & -t_{l, q}^+ \cr  -t_{l, q}^- & t_{l, q}^+}\right) e^{i\beta_{l, q}},
   \quad
    \hat T^-_{l, q} = \frac{1}{2}\left(\matrix{
    t_{l, q}^+ & t_{l, q}^- \cr t_{l, q}^+ & t_{l, q}^-}\right) e^{-i\beta_{l, q}},
\eeq
with $t_{l, q}^\pm = \cos(\alpha_{l, q})\pm \sin(\alpha_{l, q})$.
The unitary transformation of the network Eq.~\eqref{eqn:CNN}, is then given in the form of a partial Fourier series,
\begin{equation}
\fl \hat U(\vec{x}, \vec\theta) = \prod_{l = 1}^L \hat E^l\bigotimes_{q = 1}^{Q} (\hat T_{l, q}^+ e^{i \vec \omega_{l, q} \cdot \vec x} + \hat T_{l, q}^- e^{-i \vec \omega_{l, q} \cdot \vec x}) = \sum_{j=1}^{\mathcal{N}_h} \hat C_j(\alpha, \beta) \exp{(i\vec M_j(\vec \omega) \cdot \vec x)},
\label{eqn:CNN_s}
\end{equation}
where  $\alpha = \{\alpha_{l,q} \}$, $\beta = \{\beta_{l,q} \}$, $\vec \omega = \{\vec \omega_{l,q} \}$, $\hat C_j$ are $2^Q\times 2^Q$ real matrices, and $\vec M_j(\vec\omega)$ $\in$ $\mathbb{R}^n$ results from the linear combination of the distinct frequencies $\vec\omega_{l,q}$. The product of the unitaries in each layer has been rewritten as a single summation that accounts for the total number of terms $\mathcal{N}_h$ in the partial Fourier decomposition of the network. The spectrum of the QNN has a size of $\mathcal{N}_h = 2^{LQ}$, and depends solely on the data encoding strategy, regardless of the presence of entanglement, which only affects the  values of the matrix coefficients $\hat C_j$ of the Fourier series \cite{schuld2021effect}. Additional information regarding this derivation can be found in \ref{appendix_a}.

Similarly, the output generated by the QNN can be decomposed into a partial Fourier series,
\begin{equation}
    \fl
    F(\vec{x}, \vec\theta) = \bra{0} \hat U^{\dagger}(\vec x, \vec\theta)  \hat \mathcal{ M} \hat U(\vec x, \vec\theta)\ket{0} = \sum_{j=1}^{N_h} c_j(\alpha, \beta)\exp\left(i \vec \Omega_j(\vec \omega)\cdot \vec x\right).
    \label{eqn:fourier0}
\end{equation} 
The details on how to perform this decomposition are given in \ref{appendix_b}. The number of terms in this expansion is $N_h \leq 3^{LQ}$. Equation~\eqref{eqn:fourier0} describes the family of functions that a given QNN can learn through two interrelated characteristics: the spectrum of Fourier frequencies $\vec{\Omega}_j$ $\in$ $\mathbb{R}^n$ accessible by the quantum model, and the expressivity, that refers to the number of expansion coefficients $c_j$ $\in$ $\mathbb{R}$ that a class of QNN can control independently, and defines the set of attainable functions. It is important to notice that these two characteristics are fully determined by the $N_p=LQ(n+2)$ topological hyper-parameters $\vec\theta_{l,q}$, and each set plays separate roles; $\vec\omega_{l,q}$ controls the frequency spectrum, while $\alpha_{l,q}$ and $\beta_{l,q}$, the amplitude of each Fourier term $c_j$. Crucially, the number of non-vanishing expansion coefficients $N_h$ depends on the observable $\hat{\mathcal{M}}$ used to map the function $F(\vec x, \vec \theta)\rightarrow f(\vec x)$. Note that, in general, $N_h\gg N_p$. As a result, only a subset of both $c_j$ and $\Omega_j$ may be totally tunable, while the rest are constrained.
The variational formulation of a specific machine learning problem is equivalent to finding the best fit for the Fourier decomposition. While the optimal combination depends on the specific problem to be solved –i.e., $f(\vec{x})$–, the expressivity of the network is determined by the topological parameters $N_p$ and $N_h$. As we will show, the presence of entanglement, either by a global readout or a final entangling layer,  produces a larger set $N_h$ compared to a local readout in a non-entangling QNN. This enriches the accessible functions that the QNN can access, leading to an enhancement of the network accuracy when approximating $f(\vec{x})$. 

In the upcoming sections, we determine the count of non-zero expansion coefficients, denoted as $N_h$. This calculation is conducted both with and without considering entanglement between layers, and depends on the choice of observable utilized for the readout of the output state.

\subsection{One qubit network}
For the particular case of a single qubit architecture $\mathcal{A}_0 (L,1)$, we recursively count the number of vanishing $c_j(\alpha, \beta)$. According to Eq.~\eqref{eq:harmonic1q}, $L$ layers acting on a single qubit produce the transformation
\beq
\hat v = \prod_{l=1}^L (\hat T_{l}^+ e^{i \vec \omega_{l} \cdot \vec x} + \hat T_{l}^- e^{-i \vec \omega_{l} \cdot \vec x}).
\eeq
For the particular layer architecture given by Eq.~\eqref{eqn:layer}, the number of expansion terms $N_h$ depends on the considered observable $\hat{\mathcal{M}}$, being the maximum number of achivable terms $\max(N_h) = \chi = 3^L$. As shown in \ref{appendix_c}, the cases $\hat \mathcal{M} = \hat \sigma_z$ and $\hat \mathcal{M} = \hat \sigma_y$ lead to $N_h \leq 2 \cdot \chi/3$, while measuring the observable $\hat \mathcal{M} = \hat \sigma_x$ gives $N_h  \leq \chi/3$. In general, such a layer topology is capable of reaching the maximal expressivity bound $N_h = \chi$ by measuring the operator $\hat \mathcal{M} = \hat O = u_o \mathbb{I} + u_x \hat \sigma_x + u_y \hat \sigma_y + u_z \hat \sigma_z$ along various directions defined by the coefficients $u_o, u_x, u_y, u_z \in \mathbb{R}$. Notice that using another layer topology when constructing a data re-uploaded QNN might lead to a different number of non-vanishing expansion coefficients. For instance, the architecture given by
\beq\label{eqn:2nd_architecture}
\hat{\mathcal{L}}^*_{l}(\vec x, \vec\theta_{l}) = \hat R_x(2 \alpha_{l})  \hat R_y\left(2\left(\vec \omega_{l} \cdot\vec x + \beta_{l}\right)\right),
\eeq
results in $N_h  \leq 2 \cdot \chi/3 + 1$, regardless of the observable used for the readout. Table \ref{table:t1} presents the different achievable expresivities in terms of the observable $\hat \mathcal{M}$ and encoding  
layer.

\begin{table}
\centering
\begin{tabular}{|ll|llll|}
\hline
\multicolumn{2}{|c|}{$Q = 1$}                                       & \multicolumn{4}{c|}{$\hat\mathcal{ M}$}                                                                        \\ \hline
\multicolumn{2}{|l|}{}                                              & \multicolumn{1}{l|}{$\hat \sigma_x$} & \multicolumn{1}{l|}{$\hat \sigma_y$} & \multicolumn{1}{l|}{$\hat \sigma_z$} & $\hat O$ \\ \hline
\multicolumn{1}{|l|}{\multirow{2}{*}{$N_h$}} & $\hat \mathcal{ L}$   & \multicolumn{1}{l|}{$\chi /3$}           & \multicolumn{1}{l|}{$2 \chi/3$}           & \multicolumn{1}{l|}{$2 \chi/3$}           &  $\chi$        \\ \cline{2-6} 
\multicolumn{1}{|l|}{}                       & $ \hat \mathcal{ L}^*$ & \multicolumn{1}{l|}{$2 \chi/3 + 1$}           & \multicolumn{1}{l|}{$2 \chi/3 + 1$}           & \multicolumn{1}{l|}{$2 \chi/3 + 1$}           &  $2 \chi/3  + 1$        \\ \hline
\end{tabular}
\caption{Maximum number $N_h$ of non-vanishing expansion coefficients when expanding the output of the network using one qubit ($Q=1$) into a partial Fourier series. Different observables $\hat\mathcal{{M}}$ and the encoding architectures given by Eqs. \eqref{eqn:layer} and \eqref{eqn:2nd_architecture} are considered. Here $\chi = 3^L$ is the spectrum bound. } 
\label{table:t1}
\end{table}

\subsection{Multiqubit architectures}
The previous results can easily be generalized for multiqubit $\mathcal{A}_0 (L,Q)$ architectures without entanglement by taking the outer product of the unitary acting on each qubit, i.e.,
\beq\label{eqn:U_vq}
    \hat U =  \bigotimes_{q = 1}^Q \hat v_{q},
\eeq
with $\hat v_{q} = \prod_{l=1}^L (\hat T_{l, q}^+ e^{i \vec \omega_{l, q} \cdot \vec x} + \hat T_{l, q}^- e^{-i \vec \omega_{l, q} \cdot \vec x})$. The absence of the entangling gates between layers allows an exchange of products over layers and qubits in Eq.~\eqref{eqn:CNN_s} to recover Eq.~\eqref{eqn:U_vq}.

Now, two different cases are considered depending on the election of the observable $\hat{\mathcal{M}}$. At first, we consider that the mapping $\hat{\mathcal{M}}$ of the whole system is composed of local measurements $\hat{\mathcal{M}}_q$ of the individual qubits $\hat{\mathcal{M}}=\sum_{q=1}^Q\hat{\mathcal{M}}_q$. As a result, the generated output is decomposed into
\beq
\label{global}
\fl
 F(\vec{x}, \vec\theta) =\bra{0}(\bigotimes_{q=1}^Q \hat v_{q})^\dagger (\hat\mathcal{ M}_1 +  ... + \hat\mathcal{ M}_Q) (\bigotimes_{q=1}^Q \hat v_q)\ket{0} = \sum_{q=1}^Q\bra{0} \hat v_q^\dagger \hat \mathcal{ M}_q \hat v_q\ket{0}.
\eeq
Each term $\bra{0} v_q^\dagger \hat \mathcal{ M}_q \hat v_q\ket{0}$ in the expansion produces $N_h(L, 1)$ coefficients, leading to a total of $N_h(L,Q) = Q\cdot N_h(L, 1)$ non-vanishing expansion coefficients of the partial Fourier series for the $\mathcal{A}_0 (L,Q)$ architecture. 

Alternatively, we can perform a global measurement of the network by defining the observable $\hat{\mathcal{M}} = \otimes_{q=1}^Q\hat{\mathcal{M}}_q$. In this case, 
\beq
\fl
     F(\vec{x}, \vec\theta)  = \bra{0}(\bigotimes_{q=1}^Q \hat v_q)^\dagger (\hat \mathcal{ M}_1 \otimes  \hat \mathcal{ M}_2 \otimes ... \otimes \hat \mathcal{ M}_Q) (\bigotimes_{q=1}^Q v_q)\ket{0} = \prod_{q=1}^Q\bra{0} v_q^\dagger \hat \mathcal{ M}_q \hat v_q\ket{0},
\eeq
leading to $N_h(L,Q) = N_h(L, 1)^Q$ non-vanishing expansion coefficients. The same architecture can generate more harmonics by just modifying the measured observable.


\begin{table}[b]
\centering
\begin{tabular}{|ll|lll|}
\hline
\multicolumn{2}{|c|}{$Q = 2,\hat \mathcal{ L}$}                                                                                       & \multicolumn{3}{c|}{$\hat\mathcal{ M}$}                                              \\ \hline
\multicolumn{2}{|l|}{}                                                                                               & \multicolumn{1}{l|}{$\hat \sigma_z^1 + \hat \sigma_z^2$} & \multicolumn{1}{l|}{$\hat \sigma_z^1\otimes \hat \sigma_z^2$} & $\hat O_1 \otimes \hat O_2$ \\ \hline
\multicolumn{1}{|l|}{\multirow{3}{*}{$N_h$}} & without $\hat E_2$                                                                 & \multicolumn{1}{l|}{$4\chi/ 3$}      & \multicolumn{1}{l|}{$4\chi^2/ 9$}      & $\chi^2$    \\ \cline{2-5} 
\multicolumn{1}{|l|}{}                    & $\hat E_2$ all layers                                                              & \multicolumn{1}{l|}{$5 \chi^2 /9$}      & \multicolumn{1}{l|}{$4 \chi^2 /9$}      & $\chi^2$    \\ \cline{2-5} 
\multicolumn{1}{|l|}{}                    & \begin{tabular}[c]{@{}l@{}} $\hat E_2$ last layer\end{tabular} & \multicolumn{1}{l|}{$5 \chi^2 /9$}      & \multicolumn{1}{l|}{$4\chi^2/9$}     & \multicolumn{1}{l|}{$\chi^2$}    \\ \hline
\end{tabular}
\caption{Maximum number $N_h$ of non-vanishing expansion coefficients when expanding the network's output $F(\vec x, \vec\theta)$ into a partial Fourier series using $Q=2$ qubits. Different observables $\hat \mathcal{{M}}$ are considered when using the encoding architecture. In particular, the number $N_h$ is compared with and without entanglement. Here $N_h$ is compared with the maximum number of harmonics of a single qubit network $\chi = 3^L$.}
\label{table:t2}
\end{table}

\subsection{Last layer entanglement}

As commented at the beginning of Sec. \ref{approx_pow} and further elaborated by Schuld et al. \cite{schuld2021effect}, entanglement does not extend the frequency spectrum of the unitary tranformation given by Eq.~\eqref{eqn:CNN_s}, but only the  values of the matrix coefficients $\hat C_j$. A systematic approach to analyze the full tunability of $C_j$ is still needed. However, entanglement does affect the size of the spectrum of $F(\vec x, \vec \theta)$ that involves the mapping of $f(\vec x)$ trough an observable $\hat{\mathcal{M}}$.
To gain a physical understanding of the effects of introducing entanglement exclusively in the final layer, we examine a specific network architecture with $Q=2$ and the observable $\hat{\mathcal{M}}=\hat \sigma_z^{tot}=\hat\sigma_z^1+\hat\sigma_z^2$. When the last output layer incorporates an entangling gate in the form of Eq.~\eqref{eqn:en2}, the resulting neural network is equivalent to a non-entangling $\mathcal{A}_0(L,2)$ topology being mapped by the effective observable
\beq
    \hat{\mathcal{M}}'=\hat E_2^\dagger \hat \mathcal{ M}\hat E_2 = \hat E_2^\dagger \hat \sigma_z^{tot} \hat E_2 = \hat \sigma_x^1 \otimes \hat \sigma_x^2 - \hat \sigma_y^1 \otimes \hat \sigma_y^2.
\eeq
In this case, $\hat{\mathcal{M}}'$ is a global observable and increases the number of non-vanishing expansion coefficients compared to the local $\hat{\mathcal{M}}=\hat{\sigma}_z^{tot}$  measurements. More generally, the exact number of generated expansion coefficients depends on the newly transformed observable $\hat{\mathcal{M}}'$ through $\hat{E}_2$. In Table \ref{table:t2}, we provide the values of the spectrum size of $F(\vec{x},\vec{\theta})$ generated by various mapping strategies and topologies for two-qubit QNNs. The situation becomes more complex when dealing with architectures featuring entanglement across all layers, as the factorization performed in Eq.~\eqref{global} is not permitted, but it results in similar scaling behaviors, as discussed in \ref{appendix_d}. 

\begin{figure}[t]
\centering
\includegraphics[width=0.7\textwidth]{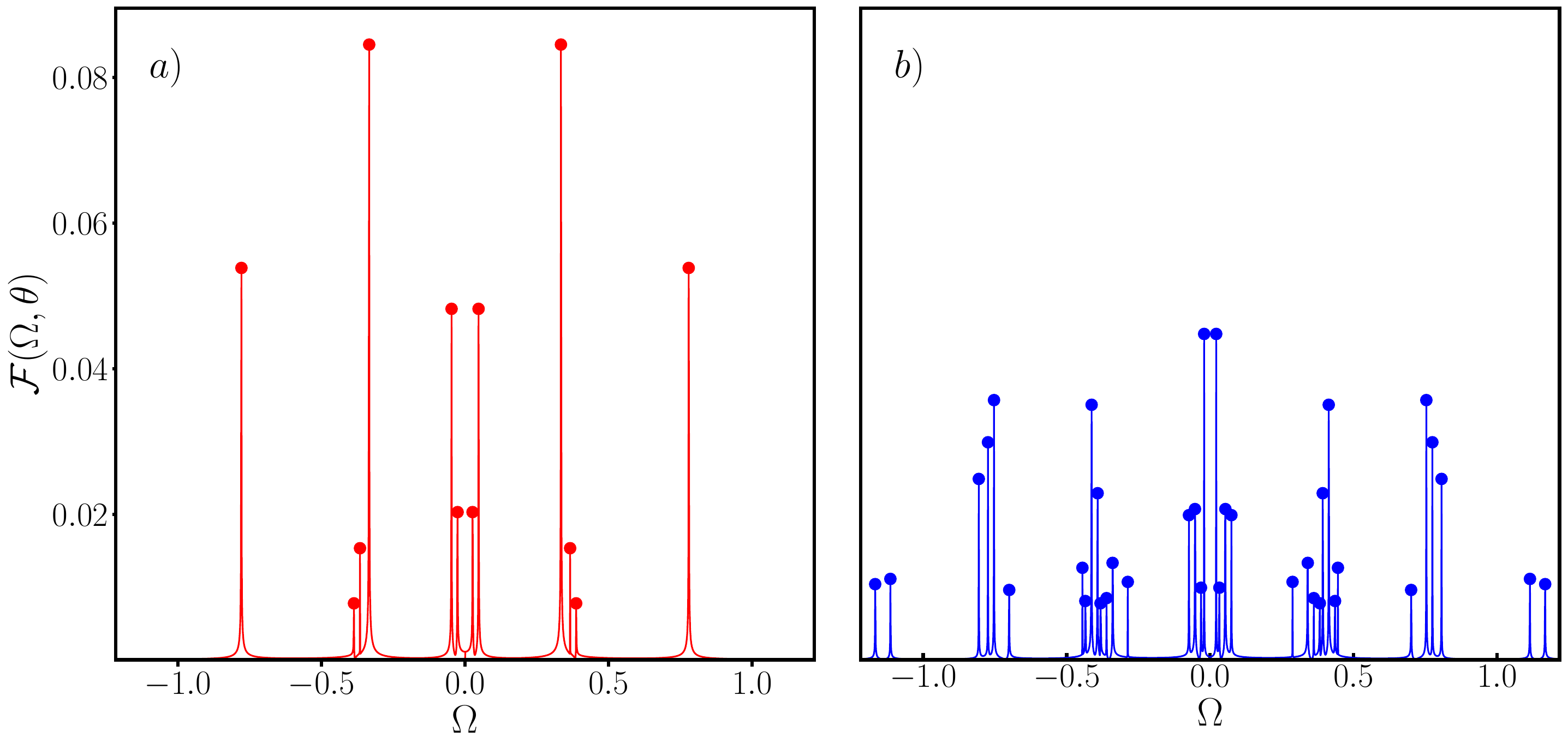}
\caption{Fourier transform of the output generated by a randomly initialized QNN showing different expansion coefficients $c_j$, by using (a)  $\mathcal{A}_0^T(2,2)$ and (b)  $\mathcal{A}_1^T(2,2)$. Sharing the same $\vec{x}$ and $\vec\theta$ values, both QNNs have the same number of variable parameters, however, the  $\mathcal{A}_1^T(2,2)$ model generates more harmonics. The measured operator is $\hat \mathcal{M} = \hat \sigma_z^{tot} = \hat \sigma_z^1 + \hat \sigma_z^2$.} 
\label{fourier_peaks}
\end{figure}

\section{Numerical results}
For simple topologies, we were able to analytically characterize the expressivity of a network by counting the number of non-vanishing expansion coefficients in the partial Fourier decomposition. However, for complex deep QNNs, this procedure becomes more intrincate. An alternative approach involves directly Fourier transforming the generated output $F(\vec x,\vec\theta)\rightarrow\mathcal{F}(\vec\Omega,\vec\theta)$, with the number of coefficients $c_j$ determined by the count of peaks in the Fourier transform. This numerical procedure allows us to extrapolate the harmonics count to deep entangling $\mathcal{A}_1 (L, Q)$ models straightforwardly. In Fig. \ref{fourier_peaks}, we plot the Fourier transform of the outputs generated by (a) $\mathcal{A}_0(2,2)$ and (b)  $\mathcal{A}_1(2,2)$ using the same encoding observable $\hat{\mathcal{M}}=\sum_{q=1}^Q\hat\sigma_z^q$ and the random but identically initialized set of parameters $\vec\theta$. The number of peaks in both the architecture without and with entanglement ($12$ and $(36)$,respectively) matches the analytical expressions in Table \ref{table:t2}. Although both $\mathcal{A}_0$ and $\mathcal{A}_1$ have the same number of tunable parameters, the inclusion of entanglement in the QNN triples the spectrum.

\begin{figure}[t]
\centering
\includegraphics[width=0.5\textwidth]{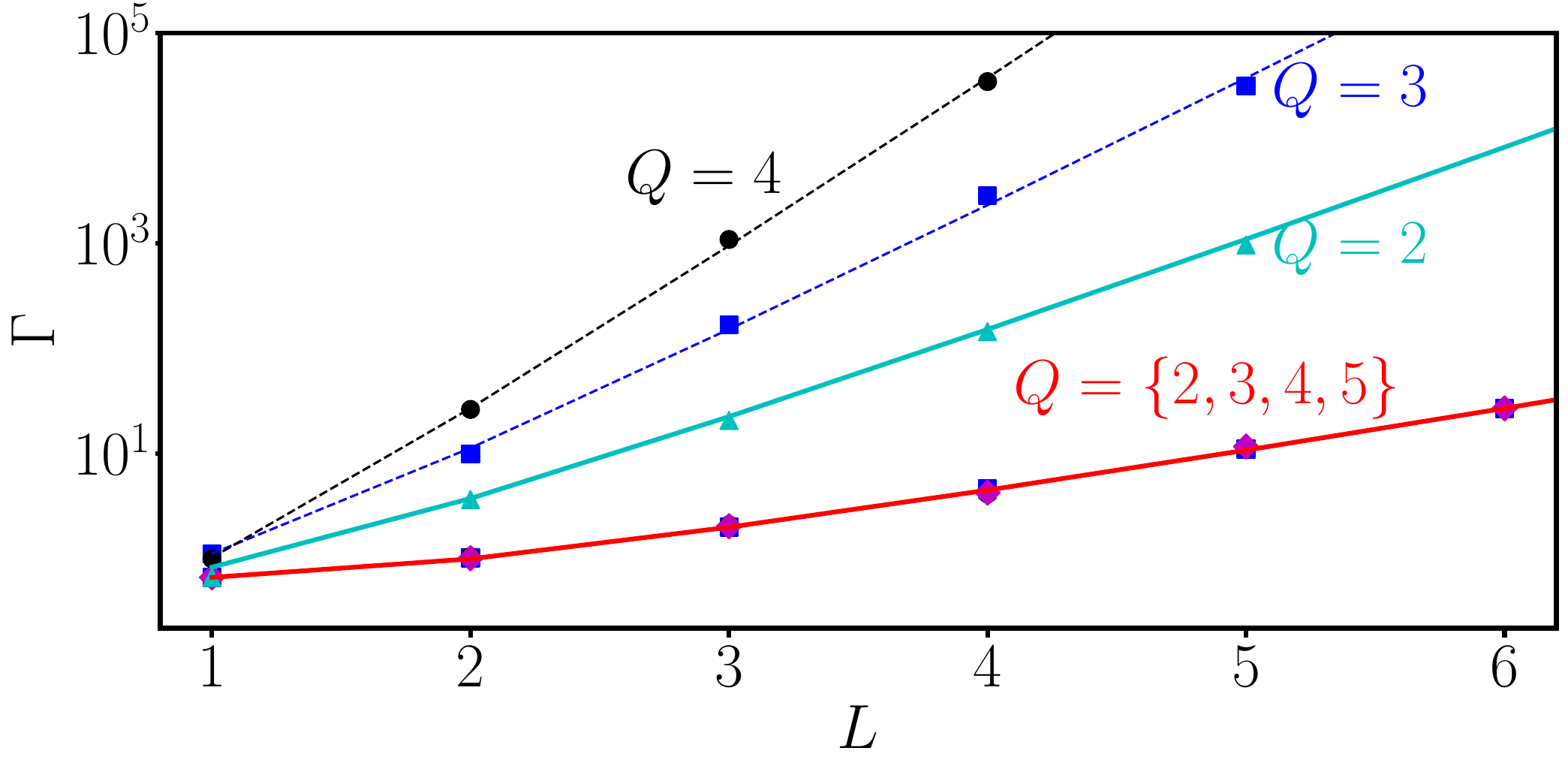}
\caption{Ratio $\Gamma$ as a function of the number of layers $L$ for different number of qubits $Q$, computed numerically through the procedure depicted in Fig. \ref{fourier_peaks} (symbols). All the $\mathcal{A}_0$ models collapse to the same ratio independently of the number of qubits (red-solid line). entangling models $\mathcal{A}_1$ show a different exponential slope for $Q=2$ (cyan triangles), $Q=3$ (blue squares), and $Q=4$ (black circles). The analytical results $\Gamma=2(3^{L-1})/[(2+n)L]$ (red-solid line) and $\Gamma=5(3^{L-1})^2/[(2+n)2L]$ (cyan-solid line) fit these trends. The dashed lines facilitate the visualization of the linear dependency. In all cases the measured operator is $\hat{\mathcal{M}} = \sum_{q=1}^Q\hat\sigma_z^{q}$. 
} 
\label{harm}
\end{figure}

To quantify the ability of the network to generate non-vanishing expansion coefficients in the partial Fourier decomposition, we define the ratio $\Gamma=N_h/N_p$. This  index quantifies the number of harmonics per adjustable parameter in the QNN, and facilitates the visualization of the QNN expressivity. Considering the encoding observable described above $\hat{\mathcal{M}} = \sum_{q=1}^Q\hat\sigma_z^{q}$, for $\mathcal{A}_0(L,Q)$ models, $N_h=Q\cdot2\chi/3$. Consequently, the ratio $\Gamma$ grows as $3^L/L$, regardless of the number of qubits, as depicted in Fig.~\ref{harm}.
In this figure, we compare this result with the $\Gamma$ parameter of $\mathcal{A}_1(L,Q)$ models, which incorporate entangling gates in all layers. While both $\mathcal{A}_0$ and $\mathcal{A}_1$ have the same number of adjustable parameters $N_p$, remarkably, the mere presence of entanglement further amplifies the exponential relationship between $\Gamma$ and $L$ as the number of qubits increases. 
%
%
%
\subsection{Teacher-student benchmark } 
In this section, we systematically evaluate the neural network approximation capabilities for real-continuous functions when considering various architectures, i.e., network depth, number of qubits, and presence or absence of entanglement in all layers. To benchmark the QNNs and minimize any potential dataset bias, we employ the teacher-student scheme \cite{Gratsea2022,zimmer2014teacher}. The teacher model (T) generates datasets by mapping random inputs to outputs, and these datasets are subsequently learned by the student models (S). To assess the performance of the student models, we employ prediction maps to qualitatively compare and visualize the similarity between T and S, see \ref{example}. For a rigorous quantitative comparison, we benchmark the average performance \cite{Hornik1991} of both models through the loss function in Eq.~\eqref{eqn:cost}.  
\subsubsection{The teacher output.}
The role of the teacher is to map the input data $\vec{x}$ through a QNN into a random output function $F(\vec{x},\vec{\theta})$ by the observable $\hat \mathcal{M} = \sum_q^Q \hat \sigma_z^{q}$, which then becomes the target $f(\vec {x})$ of S models. 
All the adjustable parameters $\vec\theta\in[-\frac{\pi}{2},\frac{\pi}{2}]$ and $K$ input values $\vec{x}\in[0,1]^n$ are randomly initialized to avoid any possible bias on the datasets. The complexity of the target function $f(\vec{x})$ can be controlled by the depth of the teacher architectures. 

\begin{figure}[b]
\centering
\includegraphics[width=0.9\textwidth]{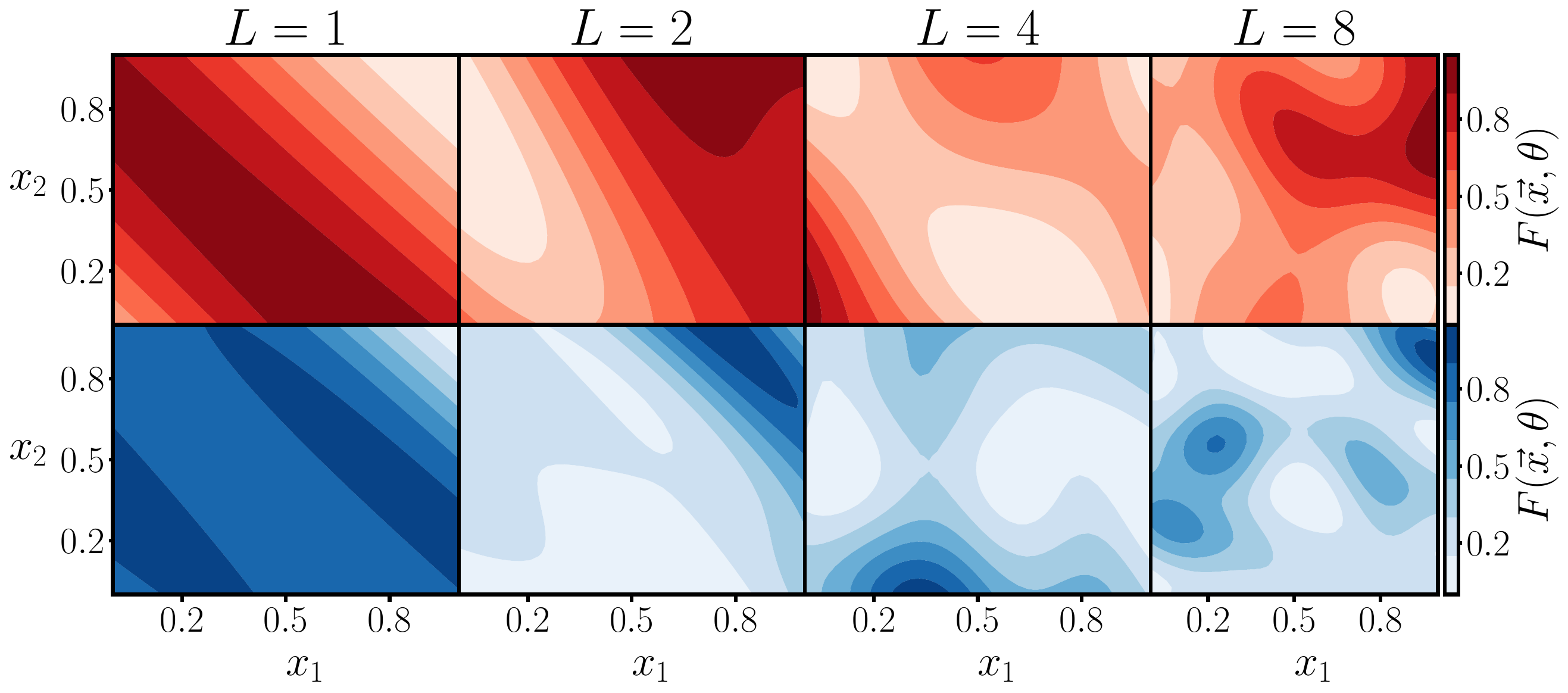}
\caption{Prediction maps generated by different T models of a QNN containing $Q=4$ qubits in different number of layers. Upper (lower) panels correspond to QNNs of  $\mathcal{A}_0^T(\mathcal{A}_1^T)$ architectures. Any pair $\{\mathcal{A}_0^T,\mathcal{A}_1^T\}$ with same $L$ is equally initialized in the network parameters $\vec\theta$ and input $\vec{x}$, but differs in the presence (absence) of entanglement.} 
\label{teacher_maps}
\end{figure}

In Fig. \ref{teacher_maps}, we show the output functions generated by several T-models containing $4$ qubits, without entanglement (upper panel) and with it (lower panel), when the input data is two-dimensional, $n=2$. Deeper architectures (left to right) provide in general a `richer' topology in $F(\vec{x},\vec{\theta})$. 

\subsubsection{The student performance.}
\begin{figure}[t!]
\centering
\includegraphics[width=0.5\textwidth]
{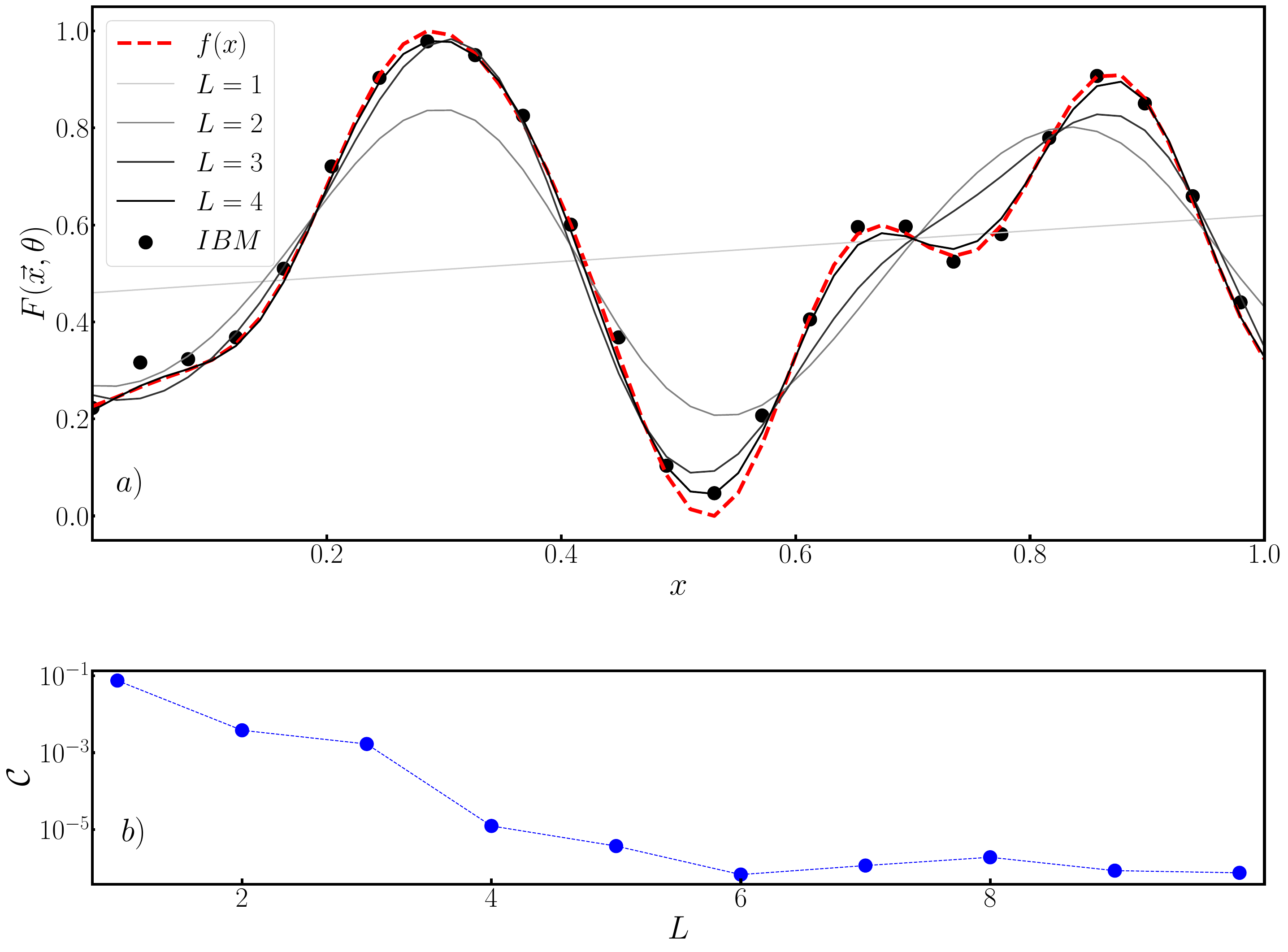}
\includegraphics[width=0.45\textwidth]
{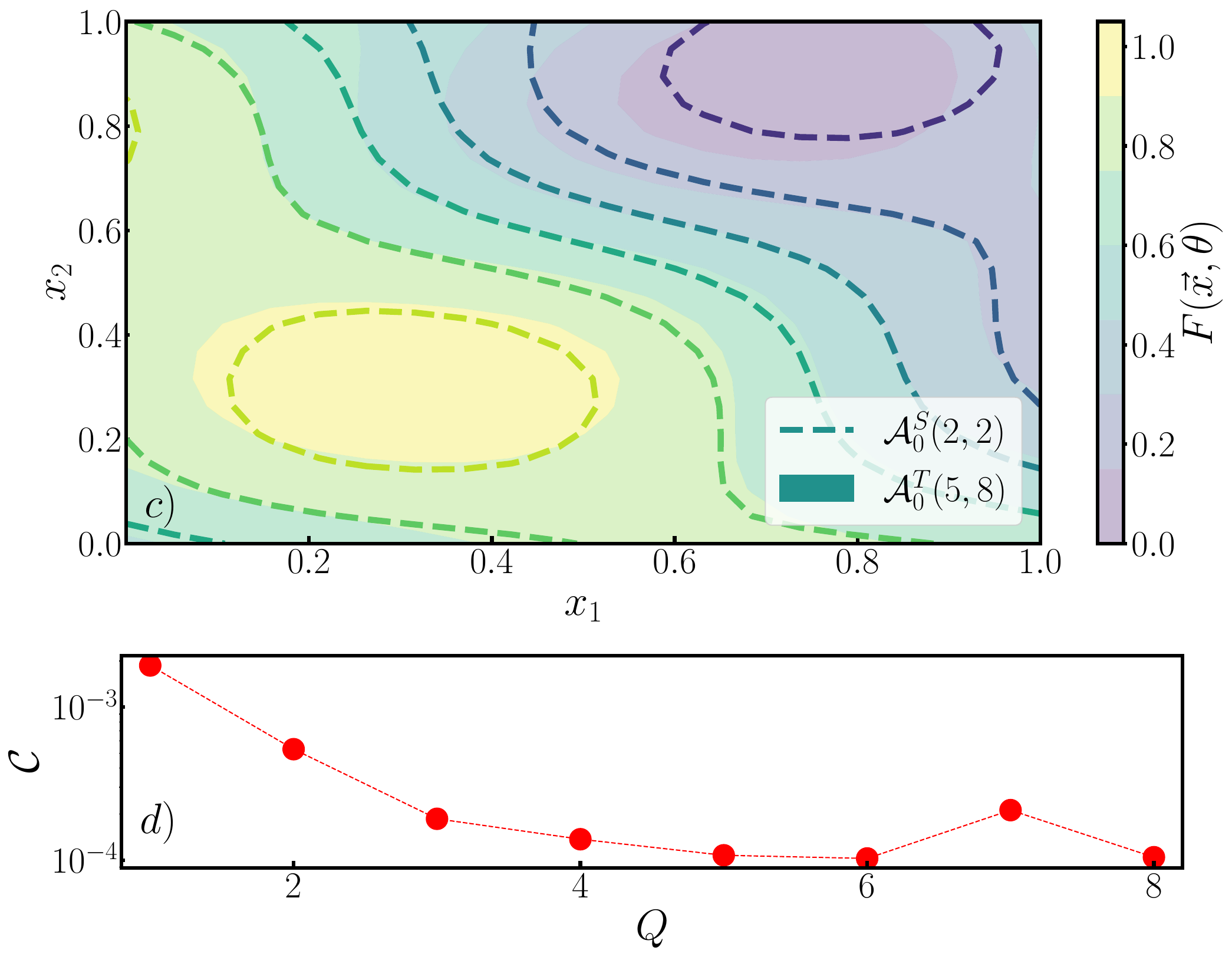}
\caption{{\itshape (a)} S-model outputs composed by a single qubit with different number of layers $L$ (solid-lines) are listed to learn the targets $f(x^k)$ function randomly generated by the T-model $\mathcal{A}^{T}_0(10,10)$ (red-dashed). A single qubit QNN $\mathcal{A}^S_0(4,1)$ run in ibmq\_belem shows its learnability (black circles). {\itshape (b)} Corresponding loss error $\mathcal{C}$ \eqref{eqn:cost} in terms of the data re-uploads $L$. 
{\itshape(c)} Output of the teacher $\mathcal{A}_0^T(5,8)$ (continuous contour plot) for a two dimensional problem $n = 2$ and the fitting result for a smaller $\mathcal{A}_0^S(2,2)$ student (dashed contour plot). {\itshape(d)} Corresponding loss error $\mathcal{C}$ for different student models  $\mathcal{A}_0^S(2,Q)$ fitting the bi-dimensional data {\itshape(c)} generated by $\mathcal{A}_0^T(5,8)$ (red-dashed line). Parameters $K=400$ and $R=10$.
}  
\label{fig:layers}
\end{figure}

First, we analyze the performance of the student as a function of the depth of the network, $L$. To this end, we focus on unidimensional ($n=1$) regression problems. We create a $\mathcal{A}^{T}_0(10,10)$ T-model, which randomly generates the output corresponding to the target function $f(x^k)$ that is depicted by the red-dashed line of Fig. \ref{fig:layers}a. Then, different $\mathcal{A}_0^S(L,1)$ S-models are trained to approximate this function. As inferred from Fig. \ref{fig:layers}b, the S-model's performance improves as the number of layers is increased, until it reaches a point of saturation, at which the minimum value of the loss function $\mathcal{C}$ becomes independent of $L$. For a fixed network topology, the particular value of $\min(\mathcal{C})$ and layer at which it saturates depend on the specific target function and input-data size. Note that a $\mathcal{A}_0^S(4,1)$ QNN programmed in ibmq\_belem (black circles of Fig. \ref{fig:layers}a) already reaches $\mathcal{C}\sim 10^{-5}$, evidencing the capabilities of a single qubit as a universal approximator \cite{PerezSalinas2021}.

%
%
%

Secondly, we study the performance of QNNs in terms of the number of qubits, $Q$, by fixing a   $\mathcal{A}_0^T(5,8)$ T-model architecture for a two-dimensional input data ($n=2$) and considering different $\mathcal{A}_0^S(2,Q)$ students. In Fig. \ref{fig:layers}c, we observe that a student with two qubits and two layers can produce an output function that closely resembles that of the teacher. In Fig. \ref{fig:layers}d, we represent the cost function as a function of the number qubits of the student. A similar trend is obtained, the approximation power of the S-models increases with the number of qubits until the loss function saturates.

Analogously to classical neural networks, where the approximation power increases with the number of hidden layers and neurons \cite{Barron1993}, the accuracy of the QNN increases with the amount of data re-uploadings (or layers) and number of qubits. However, in contrast to its classical counterpart, the QNN regression capabilities can be significantly enhanced with the incorporation of quantum resources, such as global readouts or entangling gates.

Therefore, our objective is to examine how entanglement influences the network approximation capabilities. For two-dimensional input data ($n=2$), we generate different T-models, $\mathcal{A}^T_0(L_{Max},Q)$ or $\mathcal{A}^T_1(L_{Max},Q)$. Subsequently, we evaluate the performance of S-models, $\mathcal{A}^S_0(L,Q)$ and $\mathcal{A}^S_1(L,Q)$, for both of these teacher models. The depth of the students ranges from 1 to $L_{Max}$, where $L_{Max}$ represents the number of layers in the T-models. To enlarge the benchmark, each T-model is randomly initialized $M$ times, creating multiple target functions for the students. As a result, for fixed teacher and student architectures, the QNN is evaluated $M\times R$ times, where $R$ is the number of different initializations for S-models. 

\begin{figure}[t!]
\centering
\includegraphics[width=0.9\textwidth]{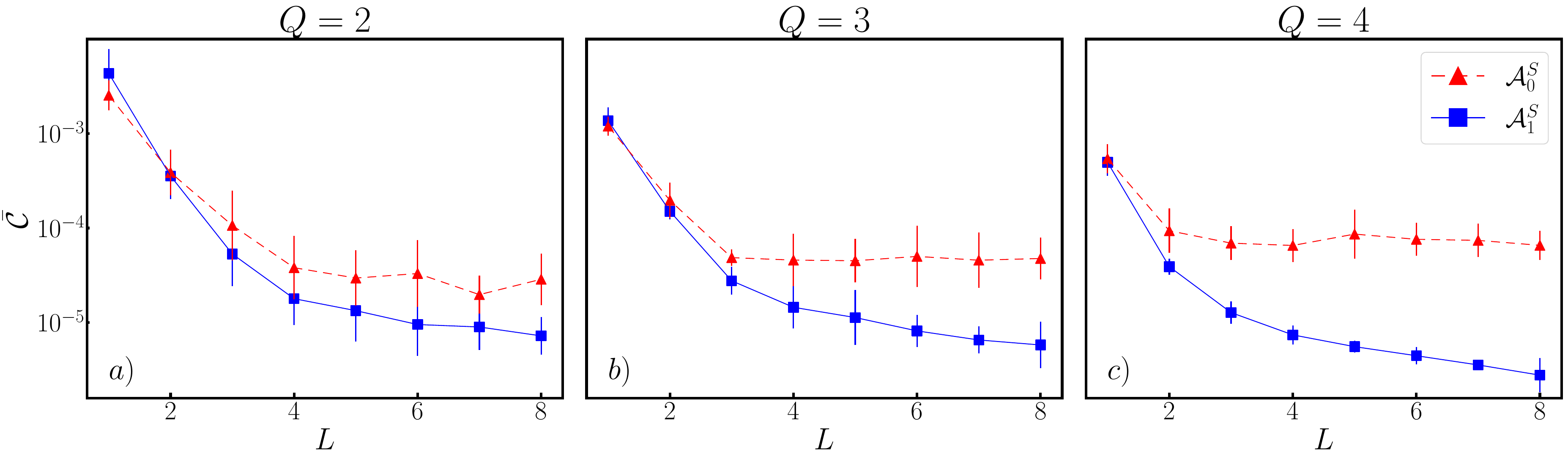}

\caption{Mean loss error  $\bar{\mathcal{C}}$ as a function of the number of layers $L$ for different student models $\mathcal{A}_0^S(L,Q)$ (red dashed-lines) and $\mathcal{A}_1^S(L,Q)$ (blue solid-lines). The target function is prepared by teacher models $\mathcal{A}_1^T(L_{Max},Q)$ containing entanglement and {\itshape (a)} $Q=2$, {\itshape (b)} $Q=3$, and {\itshape (c)} $Q=4$ qubits. Parameters: $n=2, K=400, R=5$, and $M=5$. } 
\label{ent}
\end{figure}

In Fig. \ref{ent}, we plot the student average performance $\bar{\mathcal{C}}=\sum_i^M\mathcal{C}_i/M$ over different realizations as a function of the number of layers for both the $\mathcal{A}^S_0$ and $\mathcal{A}^S_1$ models. We focus on the $\mathcal{A}_1^T$ target function, since it generates a more complex data structure due to the effect of an entangling readout. The $\mathcal{A}_0^T$ case is analyzed in \ref{noEteacher}. Nevertheless, both scenarios lead to similar conclusions. When comparing the various plots in Fig. \ref{ent}, it becomes apparent that, as the number of layers in the QNN increase, the differentiation in the performance of both $\mathcal{A}_0^S$ and $\mathcal{A}_1^S$ becomes increasingly noticeable. This observation is readily apparent in  Fig. \ref{harm}; for two QNN pairs with the same number of qubits, $\mathcal{A}_0^S$ and $\mathcal{A}_1^S$, the difference in their $\Gamma$ values grows with $L$, primarily owing to the greater number of generated harmonics in the entangling network. A similar behavior is also observed when the number of qubits is increased in the $\mathcal{A}_1^S$ topology, as it was already anticipated in Fig. \ref{harm}. For a fixed number of layers in the entangling architecture, the steepness of $\Gamma$ increases with $Q$, leading to a larger set of Fourier harmonics. However, this is not the case for $\mathcal{A}_0^S$ networks, since they all lead to the same $\Gamma$ value, regardless of the number of qubits. As a consequence, the efficiency of different $\mathcal{A}_0^S$ models saturate to a common $\bar{\mathcal{C}}$ value. This behavior is more evident in the figures presented in \ref{noEteacher}. As a result, a deep QNN involving a global mapping of $f(\vec x)$ leads to improvements of 1-2 orders of magnitude in the approximation accuracy compared to local readouts in a non-entangling network, see Fig. \ref{ent}. Finally, note that, in general, deep entangling QNNs also present higher robustness over different realizations in the predicted output, as inferred from the smaller vertical lines of Figs. \ref{ent} and \ref{fig:ex} showing the standard deviation $\Delta\mathcal{C}=\sqrt{\sum_i^M(\mathcal{C}_i-\bar{\mathcal{C}})^2/M}$.

%
%
%
%
%
%
%

%
The expressive power of the QNN primarily depends on its topological parameters. However, when considering regression tasks, other factors such the QNN readout $\hat{\mathcal{M}}$ come into play. Generally, a larger \(\Gamma\) parameter tends to result in a smaller average error \(\bar{\mathcal{C}}\), providing a higher accuracy when interpolating $f(\vec x)$, except in two distinct scenarios. Firstly, for small QNNs (as depicted in Fig. \ref{ent}a), or when dealing with simple \(f(\vec{x})\) structures (as shown in Fig. \ref{fig:ex}), there can be an overestimation of \(f(\vec{x})\). The additional harmonics introduced by \(\mathcal{A}_1^S(L,Q)\) can lead to a less accurate estimation compared to the non-entangling model \(\mathcal{A}_0^S(L,Q)\), resulting in a decrease of the network accuracy. Secondly, even though the expressive power  grows exponentially with the number of layers, we observe a saturation in the accuracy of deep QNNs. Both phenomena can be attributed to the fact that the $N_h$ Fourier coefficients have limited tunability due to the finite number $N_p=LQ(n+2)$ of adjustable parameters.

\section*{Outlook} 
\label{outlook}

We have analyzed the expressivity of a universal data re-uploading QNN and systematically benchmarked its performance in regression tasks, showing the relevance of the observable that maps the network generated output and the target function. We show that the presence of entanglement, produced by a global readout of the QNN or by the incorporation of a final entangling layer, leads to the maximal expressivity of the network. As a result, the approximation capabilities improve with respect to those based on local readouts of the output qubits.

This work paves the way to several prospective research lines. For example, {\itshape (i)} searching for alternative mapping and encoding strategies to expand the partial Fourier decomposition of the generated output, and thus enlarge the approximation efficiency of the network, such as using different building blocks for the QNN like multi-level qudits \cite{Roca-Jerat2023} or unbounded quantum systems; {\itshape (ii)} looking for applications where the designed QNN acts as a new element in a larger quantum protocol. In this context, there is the potential to design novel neural network-based quantum gates \cite{Huber2021}, which expand the quantum computing toolkit and streamline the complexity of existing algorithms. These gates can be engineered to influence the response of specific observables to input signals \cite{Torlai2018}, opening the door to new sensing schemes, as well as providing efficient representation and classification of large quantum many-body systems \cite{Gao2017, Harney2020}.

The authors thank J. J. Garc\'ia-Ripoll for fruitful discussions. We also acknowledge financial support from the Spanish Government PID2021-126694NA-C22 and PID2021-126694NB-C21, and by Comunidad de Madrid-EPUC3M14 and CAM/FEDER Project No. S2018/TCS-4342 (QUITEMAD-CM). This work is also supported by Arquimea Research Center and by Horizon Europe, Teaming for Excellence, under grant agreement No 101059999, project QCircle. H. E. acknowledges the FPU program (FPU20/03409). J. C. and E. T. acknowledge the Ram\'on y Cajal program (RYC2018-025197-I and RYC2020-030060-I). Y. B. acknowledges the CDTI within the Misiones 2021 program and the Ministry of Science and Innovation under the Recovery, Transformation and Resilience Plan-Next Generation EU under the project “CUCO: Quantum Computing and its Application to Strategic Industries”.

\appendix

\section{Derivation of the circuit unitary transformation \texorpdfstring{$\hat U_{L, Q} (\vec x, \vec\theta)$}.}
\label{appendix_a}

The unitary transformation in each layer can be decomposed as the sum of two harmonics $\omega_{l,q}$ and $-\omega_{l,q}$
\begin{eqnarray}
     \fl \hat{\mathcal{L}}_{l, q}(\vec x, \vec\theta_l) = R_x(2(\vec \omega_{l, q} \cdot\vec x + \beta_{l, q}))  R_y(2 \alpha_{l, q})= \nonumber \\
    \fl  =  
    \left(\matrix{
    \cos(\vec \omega_{l, q} \cdot\vec x + \beta_{l, q}) & -i\sin(\vec \omega_{l, q} \cdot\vec x + \beta_{l, q}) \cr
    -i\sin(\vec \omega_{l, q} \cdot\vec x + \beta_{l, q}) & \cos(\vec \omega_{l, q} \cdot\vec x + \beta_{l, q})
    }\right)
    \left( \matrix{
    \cos(\alpha_{l, q}) & -\sin(\alpha_{l, q})  \cr
    \sin(\alpha_{l, q}) & \cos(\alpha_{l, q}) 
    }\right)
      = \nonumber \\ \fl =  \hat T_{l, q}^+ e^{i \vec \omega_{l, q} \cdot \vec x} + \hat T_{l, q}^- e^{-i \vec \omega_{l, q} \cdot \vec x}, 
\label{eq:harmonic1q-detailed}
\end{eqnarray}
where 
\beq
    \hat T^+_{l, q} = \frac{1}{2} \left(\matrix{
    t_{l, q}^- & -t_{l, q}^+ \cr  -t_{l, q}^- & t_{l, q}^+}\right) e^{i\beta_{l, q}},
   \quad
    \hat T^-_{l, q} = \frac{1}{2}\left(\matrix{
    t_{l, q}^+ & t_{l, q}^- \cr t_{l, q}^+ & t_{l, q}^-}\right) e^{-i\beta_{l, q}},
\eeq
with $t_{l, q}^\pm = \cos(\alpha_{l, q})\pm \sin(\alpha_{l, q})$.
We use this expression to decompose a network into products of this sum. Then, we write these products as a factorization of single layers gates:
\beq
\hat U (\vec{x},\vec\theta) =\prod_{l=1}^L \bigotimes_{q = 1}^{Q} (\hat T_{l, q}^+ e^{i \vec \omega_{l, q} \cdot \vec x} + \hat T_{l, q}^- e^{-i \vec \omega_{l, q} \cdot \vec x}) = \prod_{l=1}^L \hat u_{l},
\label{eqn:s_CNN_s}
\eeq
with $\hat u_{l} = \bigotimes_{q = 1}^{Q}(\hat T_{l, q}^+ e^{i \vec \omega_{l, q} \cdot \vec x} + \hat T_{l, q}^- e^{-i \vec \omega_{l, q} \cdot \vec x})$. The harmonics generated by this transformation are combinations of $\omega_{l,q}$ with positive or negative sign. To compute the Kronecker product, we define $P=P(Q)$ as the set of all possible sign permutations of $Q$ terms. For example, for $Q = 3$ we have 
\begin{equation}
    P(3) = \{\{+,+,+\},\{+,+,-\},\{+,-,-\},\cdots,\{-,-,-\}\}.
\end{equation} 
We denote $P^k$ the $k$-th element in $P$, and $P^{k}_n,$ the $n$-th element in $P^k$ with $P^{k}_n = +$ or $-$ for $k = 1, ..., 2^N$ and $n = 1, ..., N$.

Using this notation, and defining $\hat T_{l, q}[\pm] = \hat T^{\pm}_{l, q}$, we express the action of a single layer as
\beq
\label{Uq0}
    \fl \hat u_{l}(\vec x,\vec\theta) = \sum^{2^Q}_{k=1} \left(\bigotimes_{q=1}^Q \hat T_{l,q}\left[P_q^k\right]\right) \exp{\left(i \sum_{q=1}^Q P_{q}^k \vec\omega_{l,q} \vec x\right)} = \sum_{k=1}^{2^Q} \hat B^k_{l} e^{i \vec m^k_{l} \cdot \vec x},
\eeq
with $\hat B^k_{l} = \hat B^k_{l}(\alpha,\beta) = \bigotimes_{q=1}^Q \hat T_{l, q}\left[P_{q}^k\right]$ and $\vec m^k_{l} =\vec m^k_{l}(\vec \omega) =  (\sum_{q=1}^Q P_{q}^k \vec\omega_{l,q})$. For illustration purposes, in the case $Q=3$, ignoring the $l$ subindex in $\hat T^{\pm}_{l,q}$ and $\vec \omega_{l,q}$ for clarity,
\beqa
\fl \hat u_{l}(\vec x,\vec\theta) = (\hat T_3^+e^{i\vec \omega_3} + \hat T^-_3e^{-i\vec \omega_3})\otimes(\hat T_2^+e^{i\vec \omega_2} + \hat T^-_2e^{-i\vec \omega_2})\otimes(\hat T_1^+e^{i\vec \omega_1} + \hat T^-_1e^{-i\vec \omega_1}) 
=  \nonumber \\  \hat T^+_3\otimes\hat T^+_2\otimes\hat T^+_1 e^{i(\vec \omega_3 + \vec \omega_2 + \vec \omega_1)\cdot \vec x} 
+\hat T^+_3\otimes\hat T^+_2\otimes\hat T^-_1 e^{i(\vec \omega_3 + \vec \omega_2 - \vec \omega_1)\cdot \vec x} + \nonumber \\ \hat T^+_3\otimes\hat T^-_2\otimes\hat T^-_1 e^{i(\vec \omega_3 - \vec \omega_2 - \vec \omega_1)\cdot \vec x} +\cdots +\hat T^-_3\hat\otimes \hat T^-_2\otimes\hat T^-_1 e^{i(-\vec \omega_3 - \vec \omega_2 - \vec \omega_1)\cdot \vec x}
\eeqa
Introducing entanglement on these architectures leads to a similar decomposition as Eq.~(\ref{Uq0}), replacing the expansion coefficients $\hat B^k_{l}\rightarrow\hat E^l\hat B^k_{l}$, while the frequency spectrum remains the same.

Introducing the expresssion for $\hat{u}_l$ obtained in Eq.~\eqref{Uq0} into Eq.~\eqref{eqn:s_CNN_s}, the action of a the whole transformation produced by the QNN is written as a sum of harmonic terms,
\beqa
    \hat U(\vec x,\vec\theta) = \prod_{l=1}^L \hat u_{l} = \prod_{l=1}^L \sum_{k=1}^{2^Q}  \hat B^k_{l} e^{i \vec m^k_{l}\cdot x} = \nonumber \\ \sum_{j=1}^{(2^Q)^L} \left (\prod_{l=1}^L \hat B_{l}^{W^{j}_l} \right)\exp{\left(i\sum_{l=1}^L \vec m_{l}^{W^{j}_l} \cdot \vec{x}\right )} = \sum_{j=1}^{2^{L Q}} \hat C_j e^{i\vec M_j \cdot \vec x},
    \label{unit_decomp}
\eeqa
where we have defined the expansion coefficients $\hat C_j = \hat C_j(\alpha, \beta) = \prod_{l=1}^L \hat B_{l}^{W^{j}_l}$, the Fourier frquencies $\vec M_j = \vec M_j(\vec \omega) = \sum_l \vec m_{l}^{W^{j}_l}$ and the indices $W^{j}_l$ are defined in a similar fashion as $P$; we take $W(L,Q)$ as the set of ordered combinations of $L$ elements which take values in $\{1, 2, ..., 2^Q \}$. For $L = 3$ and $Q = 2$,
\beq
    W(3, 2) = \{\{1, 1, 1\}, \{1, 1, 2\},\{1, 2, 1\},...,\{4,4, 3\},\{4, 4,4\}\}.
\eeq
We denote with $W^j$ the $j-th$ element in $W$, and  $W^{j}_l$ the $l$-th element in $W^j$. For the previous example $W^{2} = \{1,1,2\}$ and $W^{2}_1 = 1$. Alternatively, defining the element $X_{l, q}^j = P_q^{W^j_l}$, which takes the values $\pm$, we can write $\hat C_j = \prod_{l=1}^{L}  \bigotimes_{q=1}^{Q} \hat T_{l,q}[X_{l, q}^j]$ and $\vec M_j(\vec \omega) = \sum_{l=1}^{L} \sum_{q=1}^{Q} X_{l, q}^j \vec \omega_{l,q}$. From the expansion in Eq.~\eqref{unit_decomp}, we deduce that the total number of harmonics in the expansion of the unitary $\hat U_{L, Q}(\vec x,\vec\theta)$ is $\mathcal{N}_p = 2^{LQ}$.

\section{Derivation of the output of the network}

\label{appendix_b}

The mapping of the QNN into the output function is obtained performing a measurement in the evolved system according to Eq.~(\ref{eqn:map}), 

\beqa
    \fl F(\vec x, \vec\theta) = \bra{0} \hat U^{\dagger}  \hat \mathcal{ M} \hat U\ket{0} = \nonumber\\
    \fl \sum_{i=1}^{2^{LQ}}\sum_{j=1}^{2^{LQ}} \bra{0} \hat C^{\dagger}_{i} \hat \mathcal{ M} \hat C_j \ket{0} e^{i(\vec M_j - \vec M_i)\cdot \vec x} =  \sum_{i=1}^{2^{LQ}}\sum_{j=1}^{2^{LQ}}  \bra{0} \hat C^{\dagger}_{i} \hat \mathcal{ M} \hat C_j \ket{0} e^{i \vec \Omega_{j,i} \cdot \vec x}.
     \label{eqn:doublesum}
\eeqa
We want to write the double sum in a single sum in order to count the total number of harmonics in the expansion. Some frequencies may appear more than once, meaning that $\vec M_j - \vec M_i= \vec M_k - \vec M_n$ for some combination of $i,j\neq k,n$. We can group together the $2^{2LQ}$ terms arising from the double sum of Eq.~\eqref{eqn:doublesum} and analyze all the possible values that $\vec M_j(\vec\omega) - \vec M_i(\vec\omega)$ can yield,
\beq
    \vec \Omega_{j,i} = \vec M_{j} - \vec M_{i} = \sum_{l=1}^L \sum_{q=1}^Q (X_{l, q}^j - X_{l, q}^i)\vec \omega_{l,q}
    \label{eqn:sum_frecuencies}.
\eeq
Since $X_{l, q}^j = \{ 1, -1\}$, the difference $ X_{l, q}^j - X_{l, q}^i$ can take the values  $\{2, 0, -2 \}$. For each frequency in Eq.~\eqref{eqn:sum_frecuencies}, there exist three possible coefficients, and since there are $LQ$ frequencies $\vec{\omega}_{l,q}$, we have $3^{LQ}$ different harmonic frequencies $\vec{\Omega}_{j,i}$. As previously, let us consider $R(D)$, with $D = LQ$, the group of the permutations $\{ 2, 0, -2 \}$ of $R$ elements, such as $R = \{\{2, 2, ..., 2\}, \{2, 2, ..., 0 \}, \{2, 2, ..., -2\}, ..., \{-2, -2, ..., -2\} \}$. In this sense, $R^k$ is the $k$-th element in $R$ and $R_d^k$ the $d$-th element in $R^k$, with $k = 1,..., 3^D$ and $d = 1, ..., D$. Therefore, we can equivalently consider 
\beq
    \vec \Omega_{j,i} = \sum_{q=1}^Q \sum_{l=1}^L(X_{l, q}^j - X_{l, q}^i)\vec \omega_{l,q} \hspace{0.5cm} \longleftrightarrow \hspace{0.5cm}
    \vec \Omega_k =  \sum_{d=1}^D R^k_d \vec \omega^*_{d}.
\eeq
Defining $\vec \omega^*_d$ through a bijection with $\vec \omega_{l, q}$, such that $\vec \omega^*_1 = \vec \omega_{1, 1}$, $\vec \omega^*_2 = \vec \omega_{1, 2}$ ... $\vec \omega^*_{D} = \vec \omega_{L,Q}$.
This allows us to rewrite Eq.~\eqref{eqn:doublesum}, as a single sum with $3^{LQ}$ terms,
\beq
\fl F(\vec x, \vec\theta) = \sum_{k=1}^{3^{LQ}} \bra{0} \hat A(R^k) \ket{0} \cdot \exp\left(i \sum_d R^{k}_d \vec \omega^*_d \cdot \vec x\right) = \sum_{k=1}^{3^{LQ}} c_k e^{i  \vec \Omega_k\cdot \vec x},
\label{eqn:s_single_sum}
\eeq
with $\hat A(R^k) = \hat A^k $ an operator that comes from the combination of $\hat C^{\dagger}_i \hat \mathcal{ M} \hat C_j$ and is associated to one of the $3^{LQ}$ frequencies. For example, the coefficient $ \hat A({\{2, 0, -2, 0\}})$ is associated with the frequency $2\vec \omega^*_1 - 2\vec \omega^*_3$ (or, equivalently, $2\vec \omega_{1,1} - 2\vec \omega_{2,1}$). Some of the coefficients $c_k = c_k(\alpha, \beta) = \bra{0} \hat A^k \ket{0}$ might be zero, and thus the sum only contains $N_h\leq 3^{LQ}$. Our goal in the next section is to determine the number of non-vanishing coefficients depending on the operator that is being used to map the QNN to the output function. 

\section{Counting non-zero expansion coefficients for QNNs using a single qubit }
\label{appendix_c}

In this section of the appendix, we start by computing the matrices $\hat A(R^k)$ from Eq.~\eqref{eqn:s_single_sum} for a QNN with just one qubit and $L$ layers. The strategy consists in adding a new layer to a circuit with $L$ layers, $\hat U_L \rightarrow \hat U_{L+1} = \hat U_{L} \hat u_l$. For a single qubit network, we simply have $\hat u_{l} =(\hat T_{l}^+ e^{i \vec \omega_{l} \cdot \vec x} + \hat T_{l}^- e^{-i \vec \omega_{l} \cdot \vec x})$. We analyze the action of adding this extra layer, 
%
\beqa
    \fl \hat U_L^{\dagger}  \hat \mathcal{ M} \hat U_L =  \sum_{k=1}^{3^{L}} \hat A_L^k e^{i \vec \Omega_k \vec x} \rightarrow \nonumber \\
    \fl \hat u_l^\dagger \hat U_L^\dagger \hat \mathcal{M} \hat U_L \hat u_l =  \sum_{k=1}^{3^L} \left[ (\hat T_{l}^+)^\dagger e^{ - i \vec \omega_{l} \cdot \vec x} + (\hat T_{l}^-)^\dagger e^{i \vec \omega_{l} \cdot \vec x}\right]  \hat A_L^k  \left[\hat T_{l}^+ e^{i \vec \omega_{l} \cdot \vec x} + \hat T_{l}^- e^{-i \vec \omega_{l} \cdot \vec x}\right ] e^{i \vec \Omega_k \cdot \vec x} =  \nonumber \\
    \fl \sum_{k=1}^{3^{L}} \bigg\{ \left[(\hat T_{l}^+)^\dagger \hat A^k_L \hat T^+_{l} + (\hat T_{l}^-)^\dagger  \hat A^k_L \hat T ^-_{l}\right]e^{i\vec \Omega_k \cdot \vec x} 
    + (\hat T_{l}^-)^\dagger  \hat A^k_L \hat T^+_{l}e^{i(\vec \Omega_k + 2\vec \omega_{l}) \cdot \vec x} + (\hat T_{l}^+)^\dagger  \hat A^k_L \hat T^-_{l}e^{i(\vec \Omega_k - 2\vec \omega_{l}) \cdot \vec x}\bigg\} =  \nonumber \\
    \fl \sum_{k=1}^{3^{L}} \sum_{a \in \{0, 2, -2\}} \mathcal{H}_{l}^{a}[\hat A^k_L] e^{i(\vec \Omega_k + a \vec \omega_{l})\cdot \vec x},
    \label{eqn:derivacion}
\eeqa
Where we have introduced the superoperators
\beqa
    &\mathcal{H}_{l}^0 [\hat O] = (\hat T^+_{l})^\dagger \hat O \hat T^+_{l} + (\hat T^-_{l})^\dagger \hat O \hat T^-_{l}, \nonumber \\
    &\mathcal{H}_{l}^{+2} [\hat O] = (\hat T^-_{l})^\dagger \hat O \hat T^+_{l},  \nonumber\\
    &\mathcal{H}_{l}^{-2} [\hat O] = (\hat T^+_{l})^\dagger \hat O \hat T^-_{l}.
    \label{eqn:superoperators}
\eeqa
From the derivation in Eq.~\eqref{eqn:derivacion}, we observe that adding a layer to a single qubit network adds three new frequencies to the Fourier spectrum: $\{\Omega_k\} \rightarrow \{\Omega_k, \Omega_k + 2\omega_l, \Omega_k - 2\omega_l\}$. The coefficients of the new network are directly obtained by applying the operators $\mathcal{H}_l^a$ to the coefficients $A^k_L$, resulting in a new set of coefficients $\{A^k_L\}\rightarrow \{A^k_{L+1}\} = \{\mathcal{H}_l^0[A^k_L], \mathcal{H}_l^{+2}[A^k_L], \mathcal{H}_l^{-2}[A^k_L]\}$.
By defining the operation of adding an element $a \in \{0, 2, -2\}$ to a list $R^k$ such as $ \{R^k, a\} = \{R^{k}_1, R^{k}_2, ..., R_L^{L}, a \} $, we can now identify the relationship between the coefficients, when considering $l = L + 1$,
\beq
    \hat A_{L+1} (\{ R^k, a \}) = \mathcal{H}_{L+1}^{a}[\hat A_{L}(R^k)]
     \label{eqn:coeficients_recursive}
\eeq
from where we can obtain any term recursively as
\beq
    \hat A^k = \hat A (R^{k}) = \mathcal{H}_{1}^{R^{k}_1}[\mathcal{H}_{2}^{R^{k}_2}[...\mathcal{H}_{L-1}^{R^{k}_{L-1}}[[\mathcal{H}^{R^{k}_{L}}_{L}[\hat \mathcal{M}]]]...]].
    \label{eqn:coeficients_recursive_simple_expanded}
\eeq

The result of Eq.~\eqref{eqn:coeficients_recursive_simple_expanded} is very powerful as it allows to study the dependence of the coefficients on the measured operator $\hat{\mathcal{M}}$ for different layer architectures and find those that are zero.
In particular, for the layer definition of Eq.~\eqref{eqn:layer}, we show how $\mathcal{H}_{l,q}^a$ acts on each operator of the spin basis,
\beqa
    & \mathcal{H}_{l}^0[\hat \sigma_x] = \sin{(2\alpha_{l})} \hat \sigma_z + \cos{(2\alpha_{l})} \hat \sigma_x, \nonumber \\ 
    & \mathcal{H}_{l}^0[\hat \sigma_y] = 0, \nonumber \\
    & \mathcal{H}_{l}^0[\hat \sigma_z] = 0, \nonumber \\
    & \mathcal{H}_{l}^0[ \mathbb{I}] = \mathbb{I},
\eeqa

\beqa
    & \mathcal{H}_{l}^2[\hat \sigma_x] = 0, \nonumber \\ 
    & \mathcal{H}_{l }^2[\hat \sigma_y] = i\frac{e^{2i\beta_{l }}}{2}[\cos(2\alpha_{l }) \hat \sigma_z - \sin(2\alpha_{l }) \hat \sigma_x - i \hat \sigma_y], \nonumber \\
    & \mathcal{H}_{l }^2[\hat \sigma_z] = \frac{e^{2i\beta_{l }}}{2}[\cos(2\alpha_{l }) \hat \sigma_z - \sin(2\alpha_{l }) \hat \sigma_x - i \hat \sigma_y], \nonumber \\
    & \mathcal{H}_{l }^2[\mathbb{I}] = 0,
\eeqa

\beqa
    & \mathcal{H}_{l }^{-2}[\hat \sigma_x] = 0, \nonumber \\ 
    & \mathcal{H}_{l }^{-2}[\hat \sigma_y] = -i\frac{e^{2i\beta_{l }}}{2}[\cos(2\alpha_{l }) \hat \sigma_z + \sin(2\alpha_{l }) \hat \sigma_x + i \hat \sigma_y], \nonumber \\
    & \mathcal{H}_{l }^{-2}[\hat \sigma_z] = \frac{e^{2i\beta_{l }}}{2}[\cos(2\alpha_{l }) \hat \sigma_z + \sin(2\alpha_{l }) \hat \sigma_x + i \hat \sigma_y], \nonumber \\
    & \mathcal{H}_{l }^{-2}[\mathbb{I}] = 0.
\eeqa

If we use the observable $\hat \mathcal{ M} = \hat \sigma_z$, then all the terms associated with sequences such that $R_L^{k} = 0$, i.e., $R^k = \{R^{k}_1, ..., R_{L-1}^k, 0 \}$, return a coefficient equal to zero due to  $\mathcal{H}_{l}^0[\hat \sigma_z] = 0$. The recursive relation gives
\beqa
\hat A(\{R^{k}_1, ..., R_{L-1}^k, 0 \}) = \mathcal{H}_{1}^{R^{k}_1}[\mathcal{H}_{2}^{R^{k}_{2}}[...\mathcal{H}_{L-1}^{R^{k}_{L-1}}[\mathcal{H}^{0}_{L}[\hat \sigma_z]]...]] \nonumber \\ = \mathcal{H}_{1}^{R^{k}_1}[\mathcal{H}_{2}^{R^{k}_{2}}[...\mathcal{H}_{L-1}^{R^{k}_{L-1}}[0]...]]= 0
\eeqa
In this case, from the $3^L$ sequences that we initially had, one third of them satisfy $R^k_L=0$. Therefore, the total number of non-zero harmonic is $N_h = 2\chi/3$, where we have defined the maximum number of harmonics as $\chi = \max(N_h(L=1)) = 3^L$. Alternatively, if we choose the observable $\hat{\mathcal{M}}=\hat\sigma_x$, only sequences with terms $R^k_L = 0$ give non-vanishing terms, reaching $\chi/3$ harmonics, since $\mathcal{H}_{l}^2[\hat \sigma_x] = \mathcal{H}_{l}^{-2}[\hat \sigma_x] = 0$. The harmonics that are obtained with $\hat \mathcal{M}=\hat\sigma_x$ are precisely those that cannot be obtained with $\hat \mathcal{M}=\hat\sigma_z$. Thus, using a general operator $\hat O = u_o \mathbb{I} + u_x \hat \sigma_x + u_y \hat \sigma_y + u_z \hat \sigma_z$, we can potentially obtain $N_h = \chi$ harmonics. 

The layer definition also plays an important role on the number of harmonics. For example, if we define the layer as
\beq
\hat{\mathcal{L}}^*_{l}(\vec x, \vec\theta_l) = R_x(2\alpha_{l})  R_y(2(\vec \omega_{l} \cdot\vec x + \beta_{l})),
\label{eqn:second_architecture}
\eeq
the new superoperators $\mathcal{H}^*$ act such that
\beq
\mathcal{H}_{l}^{*0}[\hat \sigma_y] \propto  \mathcal{H}_{l}^{*0}[\mathcal{H}_{l}^{*\pm2}[\hat O]] \propto \hat \sigma_y.
\eeq 
Therefore, we can see that any sequence ending with an array of zeros after a $\pm 2$ term, such as $R^k = \{..., 2, 0\}$ or $R^k = \{..., -2,0,0, 0\}$, will make $\hat A(R^k) \propto \hat \sigma_y $. This implies that any sequence ending with a zero necessarily result in a $\hat \sigma_y$ term, except the sequence containing all zeros. Since $\bra{0} \hat \sigma_y \ket{0} = 0$, the only non-zero coefficients are those for which $R^k_L$ takes the value $\{2,-2\}$. The total number of harmonics in this case is $N^h = 2\chi/3+1$, after including the sequence containing all zeros. Now, with the application of $\hat{\mathcal{L}}^*_{l}(\vec x, \vec\theta_l)$, the total number of harmonics is independent of the measured spin operator, which is a totally different behavior from the case previously explored.  

\section{Counting non-zero expansion coefficients for QNNs using multiple qubits}
\label{appendix_d}
We can compute the superoperators $\mathcal{H}_l$ for more qubits. When adding a new layer to an existing network with Q qubits, the number of harmonics gets multiplied by $3^Q$. Thus, the number of required superoperators scales as $3^Q$, one for each new frequency, meaning that, for more qubits, the problem becomes more complicated. However, the structure of these superoperators is predictable and allows to introduce the role of entanglement. We can follow the same procedure used to derive Eq.~\ref{eqn:derivacion}, and we find
\beq
    \fl \hat u_l^\dagger \hat U_L^\dagger \hat \mathcal{M} \hat U_L \hat u_l = \sum_{k=1}^{3^{L}} \sum_{a_1 \in \{0, 2, -2\}} \sum_{a_2 \in \{0, 2, -2\}} ... \sum_{a_Q \in \{0, 2, -2\}} \mathcal{H}_{l}^{a_1, a_2, ..., a_Q}[\hat A^k_L] e^{i(\vec \Omega_k + \sum_{q=1}^Q a_q \vec \omega_{l, q})\cdot \vec x}.
\eeq
Let us take an example with two qubits
\beqa
\mathcal{H}^{0,0}_{l}[\hat O] = \sum_{i = {\pm}}  \sum_{j = {\pm}} (\hat T_{l, 2}^i \otimes \hat T_{l, 1}^j)^\dagger (\hat E_2^l)^\dagger \hat O \hat E_2^l (\hat T_{l, 2}^i \otimes \hat T_{l, 1}^j), \nonumber \\
\mathcal{H}^{0,2}_{l}[\hat O] = \sum_{i = {\pm}}  (\hat T_{l, 2}^i \otimes \hat T_{l, 1}^-)^\dagger (\hat E_2^l)^\dagger \hat O \hat E_2^l (\hat T_{l, 2}^i \otimes \hat T_{l, 1}^+),
\nonumber \\
\mathcal{H}^{0,-2}_{l}[\hat O] = \sum_{i = {\pm}}  (\hat T_{l, 2}^i \otimes \hat T_{l, 1}^+)^\dagger (\hat E_2^l)^\dagger \hat O \hat E_2^l (\hat T_{l, 2}^i \otimes \hat T_{l, 1}^-),
\nonumber \\
\mathcal{H}^{2,0}_{l}[\hat O] = \sum_{i = {\pm}}  (\hat T_{l, 2}^- \otimes \hat T_{l, 1}^i)^\dagger (\hat E_2^l)^\dagger \hat O \hat E_2^l (\hat T_{l, 2}^+ \otimes \hat T_{l, 1}^i),
\nonumber \\
\mathcal{H}^{2,2}_{l}[\hat O] = (\hat T_{l, 2}^- \otimes \hat T_{l, 1}^-)^\dagger (\hat E_2^l)^\dagger \hat O \hat E_2^l (\hat T_{l, 2}^+ \otimes \hat T_{l, 1}^+),
\nonumber \\
\mathcal{H}^{2,-2}_{l}[\hat O] = (\hat T_{l, 2}^- \otimes \hat T_{l, 1}^+)^\dagger (\hat E_2^l)^\dagger \hat O \hat E_2^l (\hat T_{l, 2}^+ \otimes \hat T_{l, 1}^-),
\nonumber \\
\mathcal{H}^{-2,0}_{l}[\hat O] = \sum_{i = {\pm}}  (\hat T_{l, 2}^+ \otimes \hat T_{l, 1}^i)^\dagger (\hat E_2^l)^\dagger \hat O \hat E_2^l (\hat T_{l, 2}^- \otimes \hat T_{l, 1}^i),
\nonumber \\
\mathcal{H}^{-2,2}_{l}[\hat O] = (\hat T_{l, 2}^+ \otimes \hat T_{l, 1}^-)^\dagger (\hat E_2^l)^\dagger \hat O \hat E_2^l (\hat T_{l, 2}^- \otimes \hat T_{l, 1}^+),
\nonumber \\
\mathcal{H}^{-2,-2}_{l}[\hat O] = (\hat T_{l, 2}^+ \otimes \hat T_{l, 1}^+)^\dagger (\hat E_2^l)^\dagger \hat O \hat E_2^l (\hat T_{l, 2}^- \otimes \hat T_{l, 1}^-).
\label{eqn:super_2q}
\eeqa

We can build the operator
\beq
\fl \hat A(R^k) = \mathcal{H}_{1}^{R^{k}_1, R^{k}_2}[\mathcal{H}_{2, 2}^{R^{k}_3, R^{k}_4}[... \mathcal{H}_{L-1}^{R^{k}_{D-3}, R^{k}_{D-2}}
 [\mathcal{H}_{L}^{R^{k}_{D-1}, R^{k}_D} [ \hat \mathcal{  M}]]  ...] ].
 \label{eqn:2qubitcoeff}
\eeq
In particular, considering the layer architecture given in Eq.~\eqref{eqn:layer}, and introducing $\hat E_2^l$ in all layers, we measure first the operator $\hat \mathcal{M} = \hat \sigma_z^{tot} = \hat \sigma_z^{1} + \sigma_z^{2}$. In this case, $\mathcal{H}^{0,2}_{l}[\hat\sigma_z^{tot}] = \mathcal{H}^{0,-2}_{l}[\hat\sigma_z^{tot}] = \mathcal{H}^{2,0}_{l}[\hat\sigma_z^{tot}] = \mathcal{H}^{-2,0}_{l}[\hat\sigma_z^{tot}] = 0$. Any sequence ending with one and only one zero in their two last terms, such as $R^k = \{..., a_{D-3}, 0, 2 \}$  or $R^k = \{..., a_{D-3}, -2, 0 \}$, yields a harmonic with zero amplitude, and as this happens for four out of nine cases, the total number of harmonics is $N_h \leq 4\chi^2/9$. If we use the operator $\hat{\mathcal{M}} = \hat \sigma_z^{1} \otimes \hat \sigma_z^{2}$ instead, the analysis yields $N_h \leq 5\chi^2/9$ harmonics. 
The maximum scaling can be achieved by measuring a general operator $\hat{O}_1  \otimes \hat{O}_2$. 

We can generalize the expressions from Eq.~\eqref{eqn:super_2q} to multiple qubits by analyzing the three possible values of $a_q$ in $\mathcal{H}_l^{...,a_q,...}:
$:
\beqa
\fl \mathcal{H}_l^{...,0,...}[\hat O] = (... \otimes  (\hat T_{l,q}^+)^\dagger \otimes ...) (\hat E^l)^\dagger \hat O \hat E^l (... \otimes \hat T_{l,q}^+ \otimes ...) + (... \otimes (\hat T_{l,q}^-)^\dagger \otimes ...) (\hat E^l)^\dagger \hat O \hat E^l (... \otimes \hat T_{l,q}^- \otimes ...) = \nonumber \\ = \sum_{i = \pm} (... \otimes (\hat T_{l,q}^i)^\dagger \otimes ...) (\hat E^l)^\dagger \hat O \hat E^l (... \otimes \hat T_{l,q}^i \otimes ...) \nonumber \\ 
\fl \mathcal{H}_l^{...,2,...}[\hat O] = (... \otimes  (\hat T_{l,q}^-)^\dagger \otimes ...) (\hat E^l)^\dagger \hat O \hat E^l (... \otimes \hat T_{l,q}^+ \otimes ...) ,\nonumber \\
\fl \mathcal{H}_l^{...,-2,...}[\hat O] = (... \otimes  (\hat T_{l,q}^+)^\dagger \otimes ...) (\hat E^l)^\dagger \hat O \hat E^l (... \otimes \hat T_{l,q}^- \otimes ...).
\eeqa
With these definitions, we can derive similar equations to Eq.~\eqref{eqn:2qubitcoeff} and analytically investigate the structure of the coefficients $c_j$ for any architecture.
\section{The teacher output}
\label{teacher}

\subsection{A particular example}
\label{example}

\begin{figure}[t!]
\centering
\includegraphics[width=0.9\textwidth]{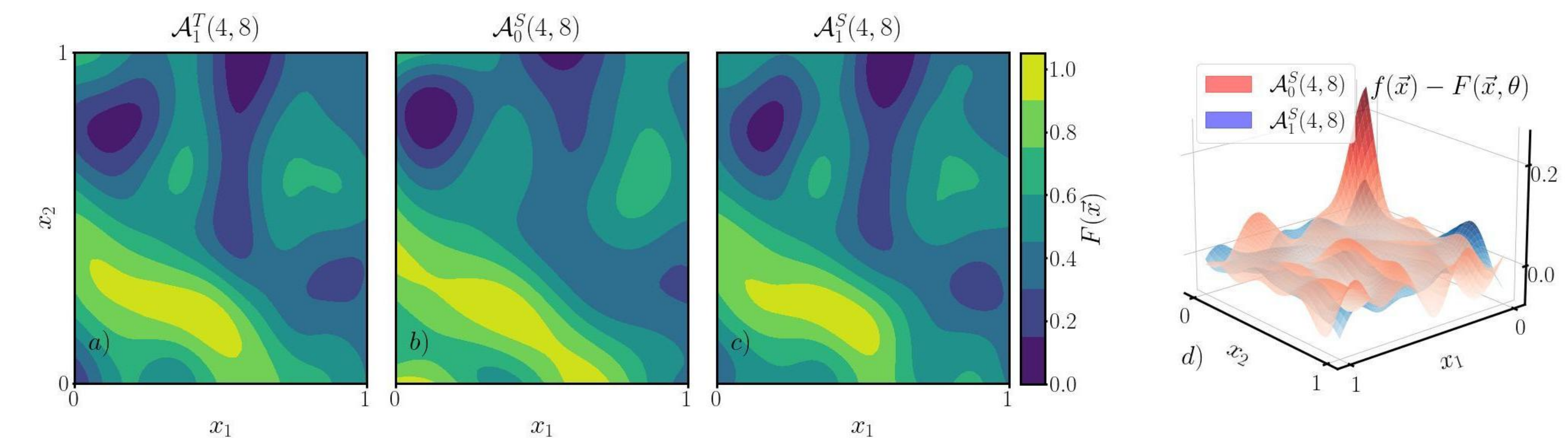}
\caption{{\itshape (a)} Bi-dimensional output generated by a teacher $\mathcal{A}_1^T(4,8)$. The contourplot corresponds to the target function to be approximated by the different student models. {\itshape (b)} Corresponging approximation generated by a student $\mathcal{A}_0^S(4,8)$ without entanglement and {\itshape (c)} by $\mathcal{A}_1^S(4,8)$ with entanglement. {\itshape (d)} Difference $f(\vec x)-F(\vec x,\vec\theta)$ for both student models. 
}
\label{fig:3d_plots}
\end{figure}

In this section, we discuss in more detail the results presented in Fig.~\ref{ent}, where different $\mathcal{A}_1^S(4, L)$ students are trained for five different $\mathcal{A}_1^T(4, 8)$ models. In Fig.~\ref{fig:3d_plots}a, we present the prediction map $F(\vec x, \vec\theta)$ of one of these $\mathcal{A}_1^T(4, 8)$ teachers. The teacher prediction map is used as the target function $f(\vec x)$ for the outputs $F(\vec x,\vec\theta)$ of the students.  For the case $L = 8$, we present in Figs. \ref{fig:3d_plots}b and \ref{fig:3d_plots}c the predictions of two S-models, $\mathcal{A}_0^S(4, 8)$ and $\mathcal{A}_1^S(4, 8)$, respectively, both initialized with the same set of parameters $\vec\theta$.

Alternatively to the values of the cost function presented in Fig.~\ref{ent}, these prediction maps allow to study the predictions locally, and understand which patterns and structures the models are able to learn. The student without entanglement is able to reproduce the overall pattern of the teacher. However, it fails when replicating smaller details, which we associate to higher frequencies. On the contrary, $\mathcal{A}_1^S$ makes a better work at approximating the teacher. Note that, in this case, both the teacher $\mathcal{A}_1^T(4,8)$ and student $\mathcal{A}_1^S(4,8)$ models are the same. However, the optimizer gets stuck in a local minimum during the minimization process, leading to small discrepancies $f(\vec x)-F(\vec x, \vec\theta)$, see Fig. \ref{fig:3d_plots}d. 

\subsection{Target function: the role of entanglement}
\label{noEteacher}

\begin{figure}[t!]
\centering
\includegraphics[width=0.9\textwidth]{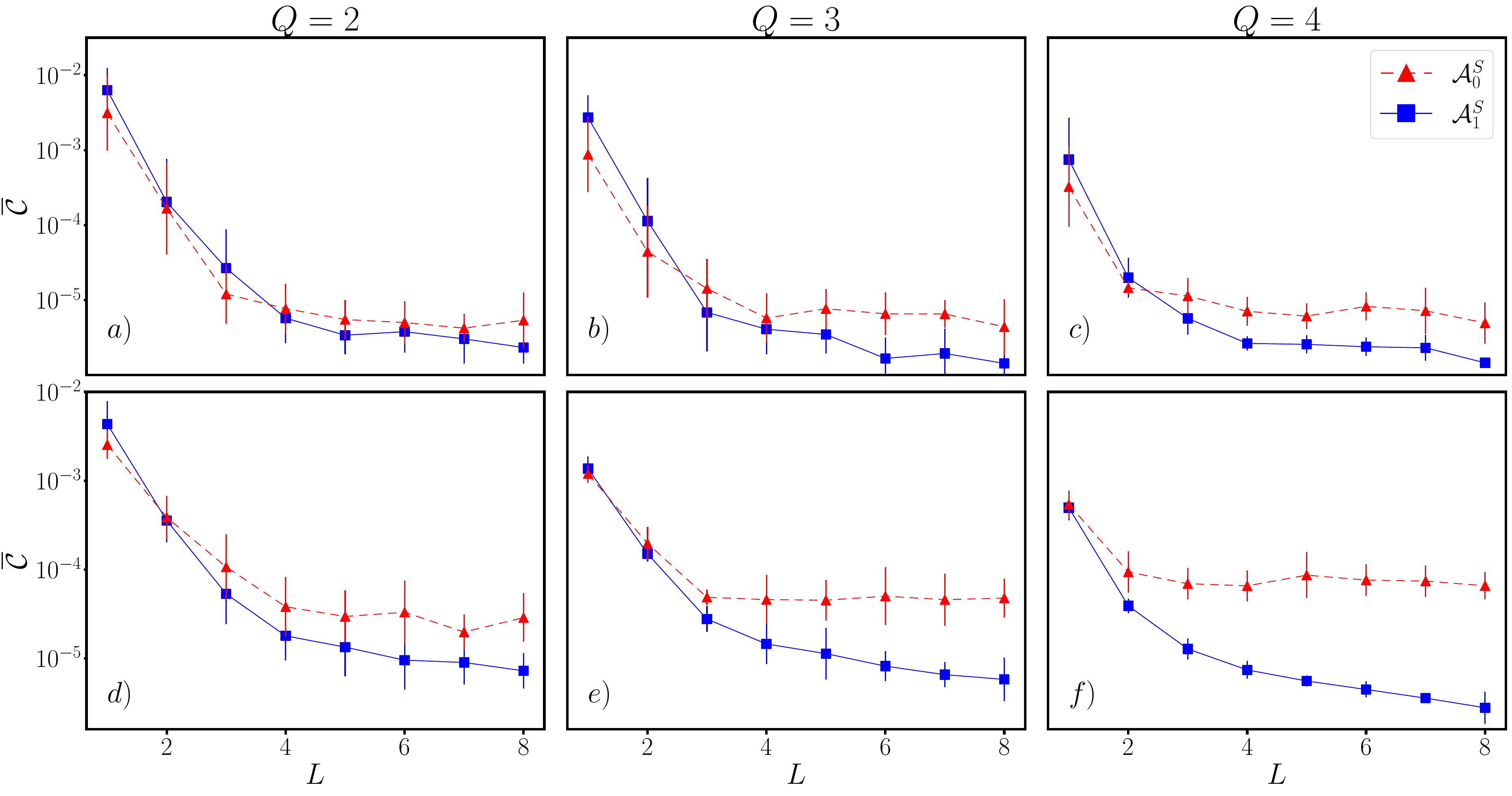}
\caption{Mean loss error  $\bar{\mathcal{C}}$ as a function of the number of layers $L$ for different student models $\mathcal{A}_0^S(L,Q)$ (red dashed-lines) and $\mathcal{A}_1^S(L,Q)$ (blue solid-lines). The target function is prepared by T-models $\mathcal{A}_0^T(L_{Max},Q)$ without and ($\mathcal{A}_1^T(L_{Max},Q)$ with) entanglement,  top (bottom) panels. From left to right, the teacher QNN contains $Q=2$, $Q=3$, and $Q=4$ qubits. Same parameters as in Fig. \ref{ent}.} 
\label{fig:ex}
\end{figure}

We have analyzed in the main text the capabilities of both $\mathcal{A}_0^S$ and $\mathcal{A}_1^S$ models to approximate a target function generated by an entangling QNN. Here, for completeness, we analyze the case of data generated by non-entangling models, $\mathcal{A}_0^T(L_{Max},Q)$, and analyze the differences with the previous case. According to  \ref{example}, the role of entanglement in the T-models is to provide a more complex target function to be approximated, and thus increases the demands of the students.
For the $\mathcal{A}_0^T(L_{Max},Q)$ models, see {\itshape (a)-(c)} in Fig. \ref{fig:ex}, we observe a small difference in the global performance of both $\mathcal{A}_0^S$ and $\mathcal{A}_1^S$. In such cases, the overall accuracy of the entangling S-model beats the non-entangling QNN, except when simple networks (low number of layers $L$) are considered. This may be attribute to an overestimation of $f(\vec{x})$ due to the non-tunability of a larger set of coefficients $c_j$ in the parial Fourier expansion (\ref{eqn:fourier0}). Note that increasing the complexity of the QNN (increasing $Q$) displaces the $\bar{\mathcal{C}}$ plots crossing point to the left, indicating that small  entangling networks $\mathcal{A}_1^S$ surpass the $\mathcal{A}_0^S$ accuracy earlier. We also observe a faster saturation of the accuracy $\bar{\mathcal{C}}$ in  $\mathcal{A}_0^S$ compared to $\mathcal{A}_1^S$. Although both QNN have the same restricted number $N_p=LQ(2+n)$ of adjustable parameters and thus accessibility to the expansion Fourier basis elements $e^{i\vec{\Omega}_j\cdot\vec x}$, the larger spectra $\Omega_j$ of $\mathcal{A}_1^S$ compared to $\mathcal{A}_0^S$ reduces the missestimation of $f(\vec x)$ by the non-tunable expansion coefficients $c_j$. As we see at the bottom panels {\itshape (d)-(f)}
\\
\\

\bibliographystyle{iopart-num}
\bibliography{apssamp}

\end{document}